\pdfoutput=1
\documentclass[preprint2]{aastex}
\usepackage{natbib}

\newcommand{\bm}[1]{\mbox{\boldmath{$#1$}}}

\shorttitle{Stabilization of Magnetically Collimated Jets}
\shortauthors{Carey et al.}
 
\begin{document}

\title{Rotational Stabilization of Magnetically Collimated Jets}

\author{C. S. Carey \altaffilmark{1} and C. R. Sovinec \altaffilmark{2}}

\altaffiltext{1}{Department of Physics, University of Wisconsin, Madison, WI 53706; cscarey@wisc.edu}
\altaffiltext{2}{Department of Engineering Physics, University of Wisconsin, Madison, WI 53706; sovinec@engr.wisc.edu}

\begin{abstract}
We investigate the launching and stability of extragalactic jets through nonlinear magnetohydrodynamic (MHD) simulation and linear eigenmode analysis. In the simulations of jet evolution, a small-scale equilibrium magnetic arcade is twisted by a differentially rotating accretion disk. These simulations produce a collimated outflow which is unstable to the current driven $m=1$ kink mode for low rotational velocities of the accretion disk relative to the Alfv\'en speed of the coronal plasma. The growth rate of the kink mode in the jet is shown to be inversely related to the rotation rate of the disk, and the jet is stable for high rotation rates. Linear MHD calculations investigate the effect of rigid rotation on the kink mode in a cylindrical plasma. These calculations show that the Coriolis force distorts the $m=1$ kink eigenmode and stabilizes it at rotation frequencies such that the rotation period is longer than a few Alfv\'en times.
\end{abstract}


\section{Introduction}
Large-scale, highly collimated energetic plasma outflows are observed in some active galactic nuclei (AGN). Many models have been proposed for the formation of these jets \citep{Ferrari:1998p1023}, but their launching, collimation, and stability remain open issues. Recent observations indicate that the magnetic field structure in AGN jets is helical in nature \citep{Asada:2005p4636, Gabuzda:2004p952, Marscher:2008p1575}. This suggests that magnetic fields play a strong role in the collimation of AGN jets, as was proposed by Blandford and Payne \citep{Blandford:1982p1468}, and that one can use a magnetohydrodynamic (MHD) model to describe their formation and evolution. However, both theory and laboratory experiments show that helical MHD equilibria can be unstable to current-driven kink modes. Understanding the effect of the kink mode on jet morphology is therefore critical to understanding their evolution. Here, we describe a computational MHD study of the stability of plasma jets relative to the kink mode and the effect that jet rotation has on the stability properties. 

Many of the earlier computational efforts to model extragalactic jets concentrate on two-dimensional MHD models in which the accretion disk is treated as a boundary condition \citep{Romanova:1997p179, Ouyed:1997p345, Ustyugova:2000p258}. Even though each of the studies cited uses a different initial magnetic field, they all observe the formation of a steady outflow. More recent three-dimensional MHD simulations study the stability of the jet far from the galactic nucleus \citep{Nakamura:2001p2683}. These calculations inject flow and torsional Alfv\'en waves into an MHD equilibrium and show that wiggled structures form in the jet due to the kink mode. Similar calculations, which consider a more realistic atmosphere into which the jet expands, also examine the effect of the kink mode on the jet \citep{Nakamura:2004p687}. These calculations show that rapid rotation of the jet can have a stabilizing effect. The study discussed here aims to further examine the effect of equilibrium rotation on the stability of an expanding jet. 

The effect of equilibrium flow on current-driven MHD instabilities has been investigated both in theory and laboratory experiments. Linear MHD calculations show that a sheared axial flow has a stabilizing effect on the kink mode, while a uniform axial flow has no effect on the growth of the instability \citep{Shumlak:1995p2833}. This effect was confirmed experimentally \citep{Shumlak:2003p1156}. Later theoretical work studied the effect of sheared helical flow on the kink mode and showed that the sheared azimuthal flow stabilizes the mode by creating a phase shift in the plasma eigenfunctions \citep{Wanex:2005p1603, Wanex:2007p2684}.

The work discussed here extends the two-dimensional simulations of jet launching \citep{Romanova:1997p179, Ouyed:1997p345, Ustyugova:2000p258} to three-dimensions via nonlinear MHD calculations and considers the effect of jet rotation on the current-driven kink mode. By scanning the rotation of the disk, we scan jet rotation, and similar to previous results \citep{Nakamura:2004p687}, the rotation of the jet is observed to stabilize the column. To better understand the stabilizing mechanism of the rotation, we perform linear MHD analysis for a simple cylindrical plasma equilibrium with rigid rotation. These calculations show that the Coriolis force stabilizes the non-resonant kink by distorting the eigenmode. 

The paper is organized as follows. Section \ref{nonlinear_jet} discusses the results of nonlinear simulations of extragalactic jet launching and evolution. The stability with regard to the kink mode is shown to depend on the rotational velocity of the accretion disk relative to the Alfv\'en speed of the initial magnetic arcade. Motivated by this result, Section \ref{livc} examines the linear stability of the kink mode in a cylindrical equilibrium with rigid rotation via initial-value MHD calculations. The results show that rigid rotation provides a stabilizing effect. In Section \ref{levc}, ideal MHD eigenvalue calculations are used to confirm the results of Section \ref{livc} and to examine the effect of equilibrium rigid rotation on the unstable range of axial wave numbers. We also examine the physical mechanism of rotational stabilization using the eigenvalue calculations in Section \ref{levc} and show that the Coriolis force stabilizes the kink mode. Discussion of the results and conclusions are given in Section \ref{conclusions}.

\section{Nonlinear Calculations of Jet Propogation}
\label{nonlinear_jet}

To investigate jet propagation, we model the expansion of a magnetic arcade due to accretion disk rotation using a non-relativistic MHD model which ignores gravitational effects. Similar to previous studies \citep{Romanova:1997p179, Ouyed:1997p345, Ustyugova:2000p258}, the accretion disk is treated as a boundary condition on the computational domain. The simulation is initialized with axisymmetric vacuum magnetic field that is tied to the disk and has zero net magnetic flux through the disk. Thus, both ends of all magnetic field lines are anchored to the accretion disk. The differential rotation of the accretion disk, which rotates with a Keplerian velicity profile, injects magnetic helicity and magnetic pressure into the magnetic field, causing it to coil and expand. The coiled magnetic field produces a hoop stress on the plasma that collimates it on the central axis. The effect of jet rotation on the stability of the column is explored by varying the rotation rate of the accretion disk in individual simulations. 

We numerically evolve the visco-resitive nonrelativistic MHD equations,

\begin{equation}
\frac{\partial n}{\partial t} + \bm \nabla \bm \cdot (n \; \mathbf v) = \bm \nabla \bm \cdot D \bm \nabla n
\label{mhd1}
\end{equation}

\begin{equation}
\frac{\partial \mathbf B}{\partial t} = \bm \nabla \bm \times (\mathbf v \bm \times \mathbf B) -  \bm \nabla \bm \times \frac{\eta}{\mu_o} (\bm \nabla \bm \times \mathbf B)
\label{mhd2}
\end{equation}

\begin{eqnarray}
\rho \frac{\partial \mathbf v}{\partial t} + \rho \; (\mathbf v \bm \cdot \bm \nabla \mathbf v) = & \frac{1}{\mu_o} (\bm \nabla \bm \times \mathbf B) \bm \times \mathbf B - \bm \nabla p \nonumber \\
& + \bm \nabla \bm \cdot \nu \; \rho \; \bm \nabla \mathbf v
\label{mhd3}
\end{eqnarray}

\begin{eqnarray}
\frac{n}{\gamma - 1} (\frac{\partial k_B T}{\partial t} + (\mathbf v \bm \cdot \bm \nabla) \; k_B T) = -\frac{1}{2} p \; \bm \nabla \bm \cdot \mathbf v \nonumber \\ 
+ \bm \nabla \bm \cdot n K \bm \nabla k_B T
\label{mhd4}
\end{eqnarray}

\noindent
where $n$ is the particle density, $\mathbf B$ is the magnetic field, $\mathbf v$ is the flow velocity, $p$ is the thermal pressure, $T$ is the ion and electron temperature, $K$ is the isotropic thermal diffusivity, $\nu$ is the viscosity, $\eta$ is the resistivity, and $\gamma$ is the ratio of the specific heats chosen such that $\gamma = 5/3$. The particle density, $n$, is related to the mass density, $\rho$, by a factor of the ion mass. The pressure and temperature are related by the ideal gas relation, assuming that the electrons and ions have the same temperature, $p = 2 n k_B T$. There is an extra term added to the right hand side of the continuity equation (Eq. \ref{mhd1}), given by $\bm \nabla \bm \cdot D \bm \nabla n$. This diffusive term is added for numerical smoothing and the diffusivity coefficient, $D$, is generally chosen to be small. The thermal diffusivity, $K$, is chosen to be $100$ times the electromagnetic diffusivity. The effect of the gravitational force due to the massive galactic central object has been ignored, so gravity does not appear in the momentum equation (Eq. \ref{mhd3}).

The MHD equations are evolved in time using the NIMROD code \citep{Sovinec:2004p1473}. NIMROD is well benchmarked and has been used to model a wide array of plasma experiments \citep{Sovinec:2003p4098} and magnetospheric physics \citep{Zhu:2006p4221}. A cylindrical computational domain with a cylindrical coordinate system given by $(r,\theta,z)$ is used. The spatial discretization scheme combines two numerical methods. A mesh of high-order finite elements is used in the poloidal ($r$-$z$) plane, where the degree of the polynomial basis functions is chosen by the user, and the azimuthal ($\theta$) direction is represented with finite Fourier series. The parameter $m$ is used to identify Fourier components in the azimuthal direction. Convergence studies show that a resolution of $0 \leq m \leq 5$, is sufficient for the dynamics of the expanding jet. Using logarithmic packing of the poloidal mesh on the central axis and disk boundary, we can resolve the jet dynamics in a large domain using a poloidal mesh of $48$ by $48$ fifth-order elements.

Previous studies searching for steady state outflows have treated the outer boundaries of the domain with open boundary conditions, allowing kinetic and magnetic energy to flow out of the domain \citep{Romanova:1997p179, Ouyed:1997p345, Ustyugova:2000p258}. We use closed, perfectly conducting boundary conditions on the outer boundaries to avoid inward propagating wave characteristics. While this boundary condition is certainly unphysical, the outer boundaries are placed at a distance of $r = z = 100 \; r_i$, where $r_i$ is the inner radius of the accretion disk, which is far from the dynamic region of the calculation. 

The model of the accretion disk/jet system treats the accretion disk as a boundary condition at $z=0$, where a smoothed axisymmetric Keplerian velocity profile is applied to $v_\theta$:

\begin{equation}
v_\theta(r,\theta,z=0) = \frac{\sqrt{GM} \; r}{(r^2 + r_i^2)^{3/4}} \; .
\end{equation}

\noindent
The remaining components of the fluid velocity on the disk boundary at $z=0$ are chosen such that $v_r = v_z = 0$. A Dirichlet boundary condition is applied on the other, distant boundaries with $\mathbf v = 0$. On all of the boundaries, the number density is constrained to be constant. Mass is allowed to diffuse through the disk boundary (to fill in the coronal mass that is removed by the jet flow) by shaping the diffusivity parameter, $D$, in Eq. \ref{mhd1} such that it is large near the disk boundary and small in the rest of the domain. All of the boundaries are treated as perfect conductors by holding the normal component of the magnetic field constant in time.

The initial condition is a currentless coronal magnetic field, just above the accretion disk. This field is chosen such that there is zero net magnetic flux through the accretion disk boundary. The poloidal magnetic flux, $\psi$, defined by $\mathbf B = \bm \nabla \psi \bm \times \bm \nabla \theta$, is chosen to be

\begin{equation}
\psi(r, z=0) = r^2 \left[1 + \left( \frac{r}{r_i} \right)^2 \right]^{-\alpha} e^{-r^2 / r_o^2} \; 
\label{ncjl_2}
\end{equation}

\noindent
on the disk boundary, where $\alpha$ is a parameter with $ 0 < \alpha < 1$. The initial magnetic field is found in terms of the magnetic potential, $\Phi$, which satisfies Laplace's equation and is given by

\begin{equation}
\mathbf B = \bm \nabla \Phi.
\label{ncjl_3}
\end{equation}

\noindent
An analytic solution for $\Phi$ in the domain is found by solving the boundary value problem $\bm \nabla^2 \Phi = 0$, where the normal component of $\bm \nabla \Phi$ on the accretion disk boundary is specified by Eq. \ref{ncjl_2},

\begin{equation}
\frac{\partial \Phi}{\partial z} = \frac{1}{r} \frac{\partial \psi}{\partial r},
\label{ncjl_4}
\end{equation}

\noindent
and $\Phi = 0$ on all of the other boundaries. The poloidal flux on the disk boundary increases from zero at $r=0$ to a maximum value at the O-point of the magnetic field, defined to be the radius where $B_z = 0$, and then exponentially decays to zero. For the calculations discussed here, the values $\alpha = 3/4$ and $r_o =  30 \; r_i$ are used. For this choice of $\alpha$, the radius of the O-point of the magnetic field is at $15.13 \; r_i$. 

The initial number density and temperature are constant across the computational domain. Thus, the entire domain is initially filled with a plasma that is essentially unmagnetized away from the initial arcade, and the magnetized jet expands into this thermal plasma. The initial flow velocity is set to zero, and the accretion disk flow is ramped from zero at $t=0$ to a steady profile within one turn of the disk at $r = r_i$. The Keplerian flow of the disk acts to twist the coronal magnetic field, building magnetic pressure above the disk which launches the outflow. This twisting of the magnetic field also creates a strong $\theta$-component to the field, causing a hoop stress which pinches the plasma on the central axis and collimates the outflow. 
 
The results discussed here are given in units of the initial field quantities. All velocities are given in units of the Alfv\'en velocity at the origin, ${v_A}_o = B_o \; \rho_o^{-1/2}$, where $B_o$ and $\rho_o$ are the magnetic field and mass density at the origin respectively. The magnetic field is given in units of $B_o$. Time is given in units of $T_i$, the rotation period of the disk at $r = r_i$.

Four dimensionless parameters are used to describe the system. The first three are commonly used to describe plasma systems: the Lundquist number, $\textrm{S} = \tau_{R} \; \tau_{A_o}^{-1}$, where $\tau_R = \mu_o \; \pi \; r_i^2 \; \eta^{-1}$ is the resistive diffusion time across the inner radius of the accretion disk and ${\tau_A}_o = r_i \; {v_A}_o^{-1}$ is the Alfv\'enic propagation time across the inner radius of the disk based on the Alfv\'en speed at the origin; the magnetic Prandtl number, $\textrm{P}_\textrm{M} = \nu \; \mu_o \; \eta^{-1}$; and the plasma beta at the origin, $\beta = P_T \; P_B^{-1}$, where $P_T$ is the thermal pressure and $P_B$ is the magnetic pressure. The last dimensionless parameter, the drive parameter, $\hat V_D$, is defined as

\begin{equation}
\hat V_D = \frac{v_\theta(r = r_i,z=0)}{v_A(r=r_i, z=0)} \; ,
\end{equation}

\noindent
where $v_A$ is the Alfv\'en velocity. This parameter can be understood as the ratio of how fast the accretion disk twists coronal magnetic field lines to how fast the information of this twisting propagates through the corona. In order to maintain a constant resistive diffusion time relative to the rotation period of the accretion disk in different simulations, the parameter $S \bm \cdot \hat V_D$ is held constant as $\hat V_D$ is varied. Three sets of parameters are considered: P$_\textrm{M}$, $\beta$, and the product $\textrm{S} \bm \cdot \hat V_D$ are fixed at $1$, $1$, and $200 \pi$ respectively, and $\hat V_D$ is varied with the values $0.5$, $1.0$, and $4.0$. The values of $\hat V_D$ are chosen to be similar to previous studies \citep{Moll:2008p4392, Nakamura:2004p687, Ouyed:2003p337} which consider sub-Alfv\'enic disk rotation, and to extend the disk rotation to the previously unstudied super-Alfv\'enic regime.

\begin{figure}[t]
\plotone{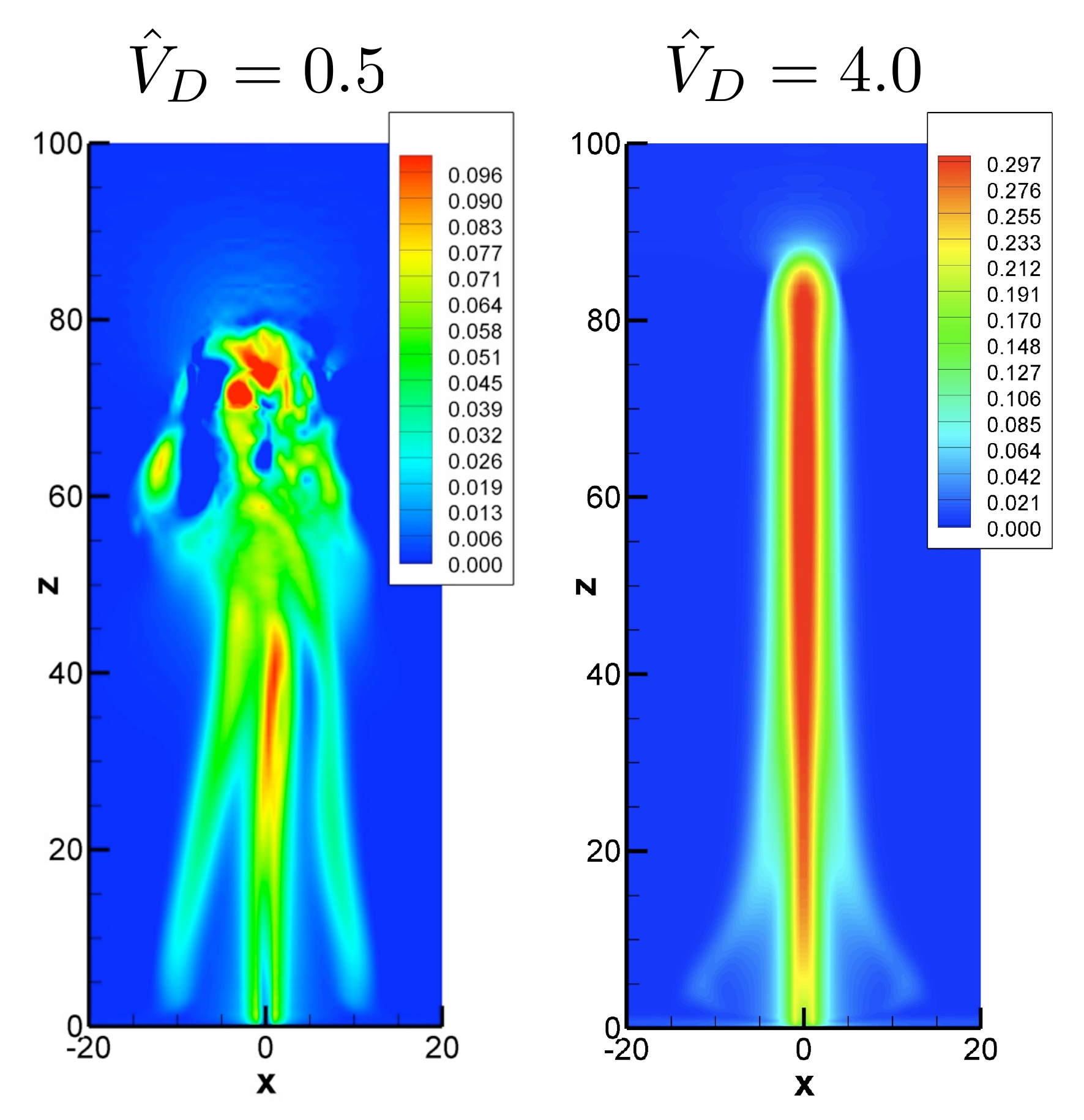}
\caption{Cross section of the $z$-component of the fluid velocity, for $\hat V_D = 0.5$ and $4.0$, at times $t = 65.6$ and $121.7 \; T_i$ respectively. Velocity is shown in units of ${v_A}_o$. Note that the domain extends to $100 \; r_i$ in radius and height.}
\label{nonlinear_jet_vz}
\end{figure}

\begin{figure}[t]
\plotone{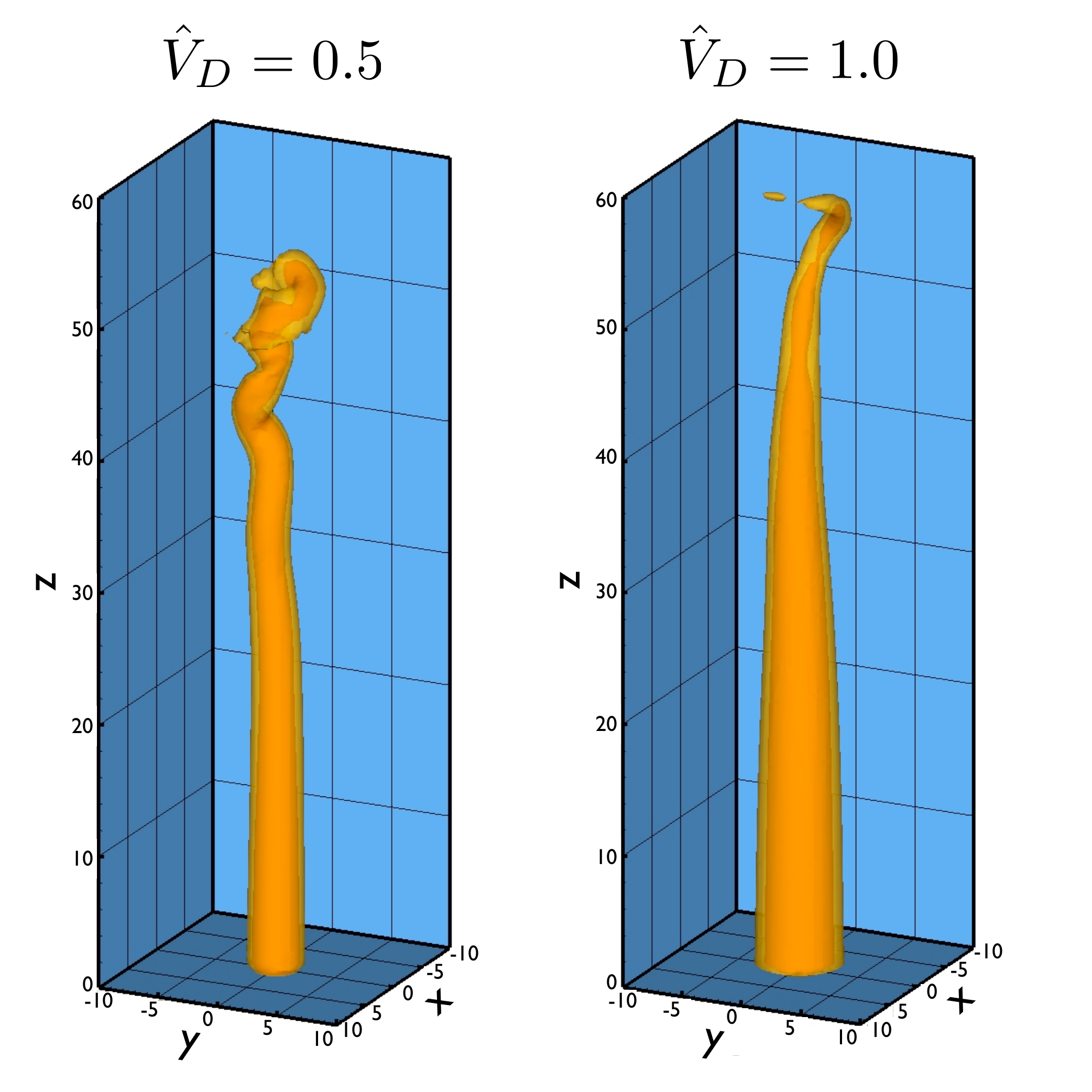}
\caption{Three-dimensional contours of the magnitude of the magnetic field ($B = 0.34 \; B_o$) for $\hat V_D = 0.5$ and $1.0$ after the $m=1$ kink mode has saturated at $t = 43 \; T_i$. Note that the domain extends to $100 \; r_i$ in radius and height.}
\label{modB}
\end{figure}

The $z$-component of the fluid velocity for the $\hat V_D = 0.5$ and $4.0$ calculations at $t = 65.6$ and $121.7 \; T_i$ respectively is shown in Fig. \ref{nonlinear_jet_vz}. While a collimated outflow is produced for both values of $\hat V_D$, non-axisymmetric structure forms in the column for the $\hat V_D = 0.5$ case, due to the presence of an MHD instability. The effect of the instability on the magnetic structure of the jet can be seen in Fig. \ref{modB}. Here the magnitude of the magnetic field is shown for $\hat V_D = 0.5$ and $1.0$ at $t = 43 \; T_i$. While the jet has expanded to a similar length for both values of $\hat V_D$ at these times, the modification of the magnetic structure is more significant for $\hat V_D = 0.5$. For both cases an $m=1$ kink mode creates a helical distortion to the magnetic structure. 

\begin{figure}[t]
\plotone{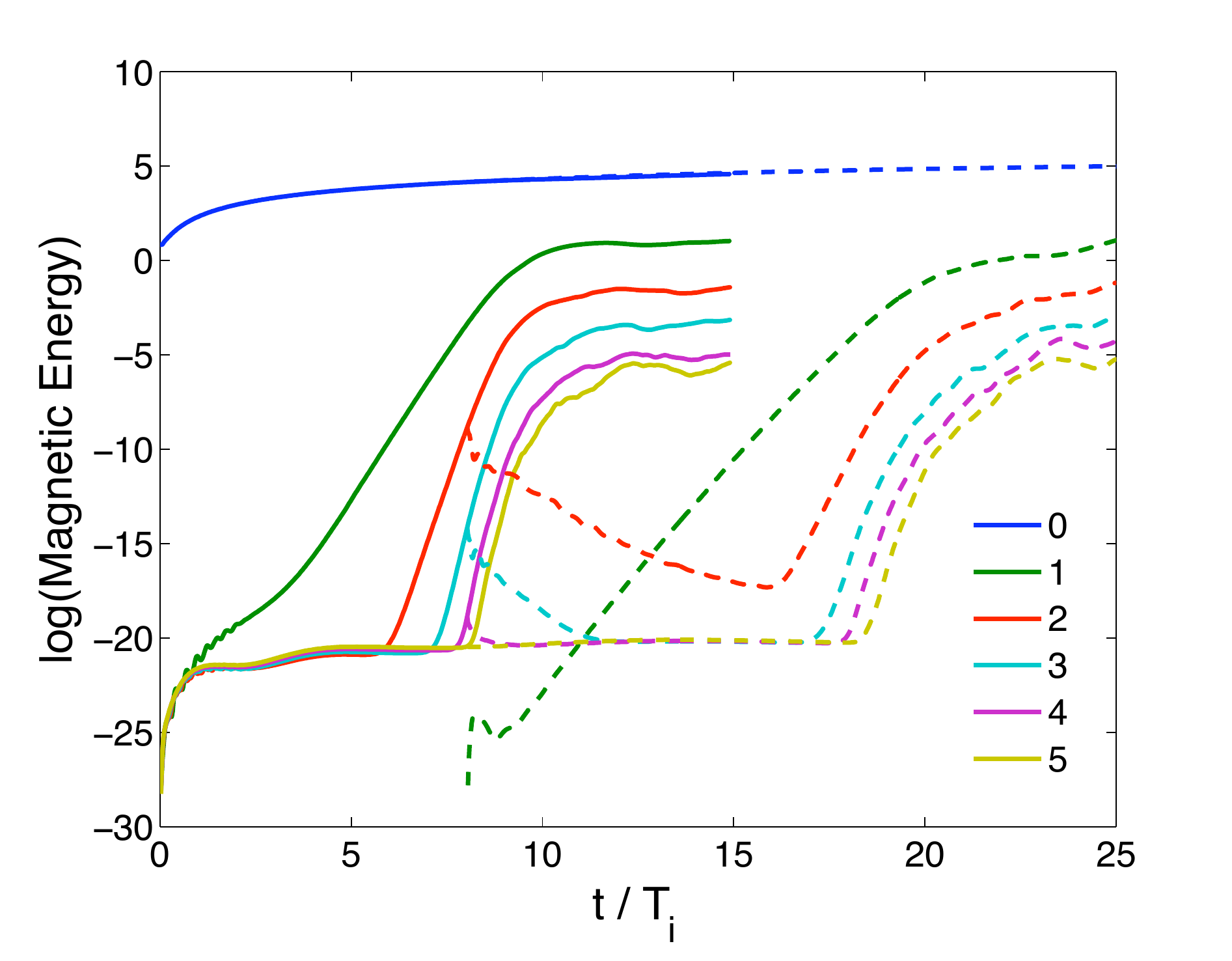}
\caption{Solid lines show the energy of individual azimuthal Fourier components as a function of time for $\hat V_D = 0.5$. The dashed lines show results from a second computation where the fields in the $m=1$ component are reset to zero at $t = 8.0 \; T_i$. The magnetic energy is normalized to $V_{dom} B_o^2 / (2 \mu_o)$, where $V_{dom}$ is the volume of the computational domain.}
\label{jet_energy_spectrum_with_m_1_removed}
\end{figure}

To confirm the source of the asymmetry in the $\hat V_D = 0.5$ simulation, we plot the energy of individual Fourier components in Fig. \ref{jet_energy_spectrum_with_m_1_removed}. The $m=1$ component is the first to become unstable, and it nonlinearly drives the $m > 1$ components when it reaches a significant level. The nonlinear drive is confirmed by artificially resetting the dependent fields in the $m=1$ component to zero during the course of a simulation. As can be seen from the dashed traces in Fig \ref{jet_energy_spectrum_with_m_1_removed}, removing the $m=1$ component causes the larger-$m$ components to decay, until the $m=1$ returns to a significant level. Thus, the $m=1$ component nonlinearly channels energy into the $m > 1$ components.

\begin{figure}[t]
\plotone{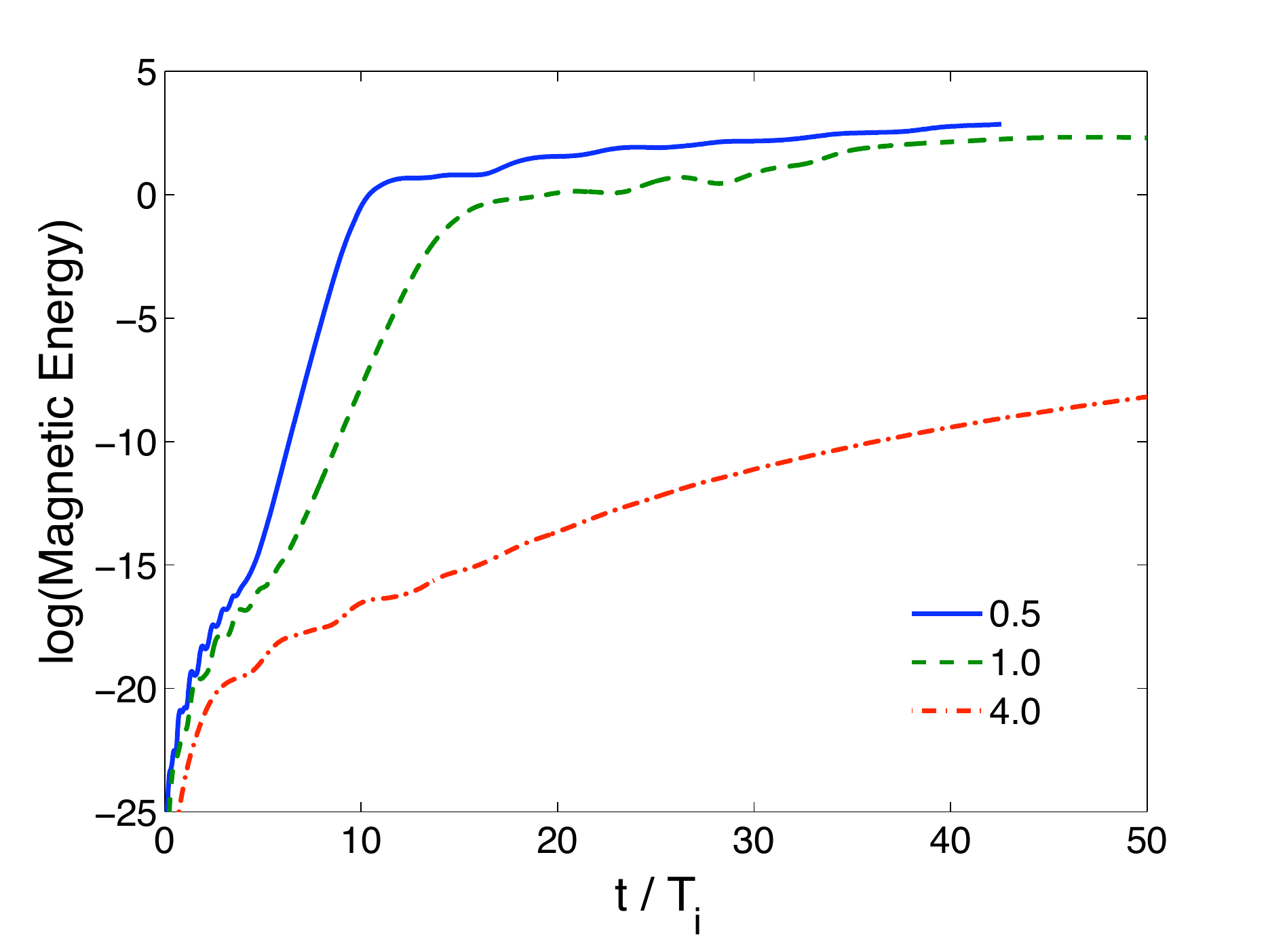}
\caption{Energy of the $m=1$ Fourier mode as a function of time for $\hat V_D = 0.5$ (solid line), $1.0$ (dashed line), and $4.0$ (dashed-dotted line). For $\hat V_D = 0.5$ and $1.0$ the jet is observed to be unstable to the $m=1$ kink mode, while for $\hat V_D = 4.0$ the jet is nearly stable. The magnetic energy is normalized to $V_{dom} B_o^2 / (2 \mu_o)$, where $V_{dom}$ is the volume of the computational domain.}
\label{m_1_energy}
\end{figure}

A plot of the magnetic energy of the $m=1$ Fourier component for all three jet simulations is shown in Fig. \ref{m_1_energy}. The jet is unstable to an $m=1$ mode for $\hat V_D = 0.5$ and $1.0$, while it remains nearly stable for $\hat V_D = 4.0$. We calculate the linear growth rate of the $m=1$ mode by making a linear fit to the magnetic energy of the $m=1$ mode when it is in the linearly growing phase. The growth rates for $\hat V_D = 0.5$, $1.0$, $4.0$ are found to be $2.98$, $1.88$, and $0.13 \; T_i^{-1}$ respectively. Thus, we see that as the accretion disk rotation increases relative to the Alfv\'en velocity of the coronal plasma, i.e. as the drive for the jet increases, the growth rate of the $m=1$ kink mode decreases.  

\section{Linear Initial Value Calculations}
\label{livc}

The stability of the kink mode for large values of $\hat V_D$ indicates that flow plays an important role in stabilizing the jet. Motivated by this observation, we perform linear initial value calculations in a simple geometry where the flow can be scanned systematically and consider both rotation and axial flow. A cylindrical domain is used with a coordinate system given by $(r, \theta, z)$. The fields are defined to be periodic in the $z$-direction, and the boundary at $r = r_a$ is treated as a perfect conductor, where $r_a$ is much smaller than the radius of the domain of the nonlinear simulations described in Sec. \ref{nonlinear_jet}. The results are given in terms of the Alfv\'en propagation time across the radius of the cylinder, $\tau_A$ = $r_a \; v_A^{-1}$, where $v_A$ is the Alfv\'en speed at $r = 0$. Here, we solve a linear version of Eqns. \ref{mhd1}-\ref{mhd4} for perturbations to MHD equilibria, with an arbitrary perturbation included in the initial velocity field. The dissipation coefficients are chosen to give a Lundquist number of $S = 1 \bm \times 10^6$ and a magnetic Prandlt number of $P_M = 1$. If an MHD equilibrium is unstable, the solution obtained will be the most unstable linear eigenmode, and the growth rate is determined from the resulting exponential growth. 

Our MHD equilibria are based on the paramagnetic pinch \citep{Bickerton:1958p2343}, which is a one-dimensional Ohmic equilibrium with uniform axial electric field. The equilibrium is characterized by the parallel current profile, $\lambda(r)$, defined as

\begin{equation}
\lambda(r) = \frac{\mu_o \mathbf{J_0(r)} \bm \cdot \mathbf B_0(r)}{B_0(r)^2} = \frac{{E_0}_z {B_0}_z(r)}{\eta B_0(r)^2},
\label{stabalization_1}
\end{equation}

\noindent
where a subscript $0$ is used to represent equilibrium fields. The profile discussed here is defined in terms of the on-axis parallel current, $\lambda_o = \lambda(r = 0)$, and the width of the equilibrium current profile decreases with increasing $\lambda_o$. Given that the $-1/2 \int \lambda \; \delta \mathbf E^{*} \cdot \delta \mathbf B \; d \mathbf x$ term is the only potentially destabilizing term in the linear ideal potential energy that is independent of $\nabla p_o$ \citep{Freidberg:1987p2544pg259}, the parallel current is related to the free magnetic energy available to drive the kink mode. Moreover, for the paramagnetic pinch, $\lambda_o$ serves as a stability parameter for the mode. The equilibrium magnetic field is found by choosing a value for $\lambda_o$ and numerically integrating Ampere's Law, $\bm \nabla \bm \times \mathbf B_0 = \mu_o \mathbf J_0$, using Eq. \ref{stabalization_1} for the parallel component of $\mathbf J_0$.

\begin{figure}[t]
\plotone{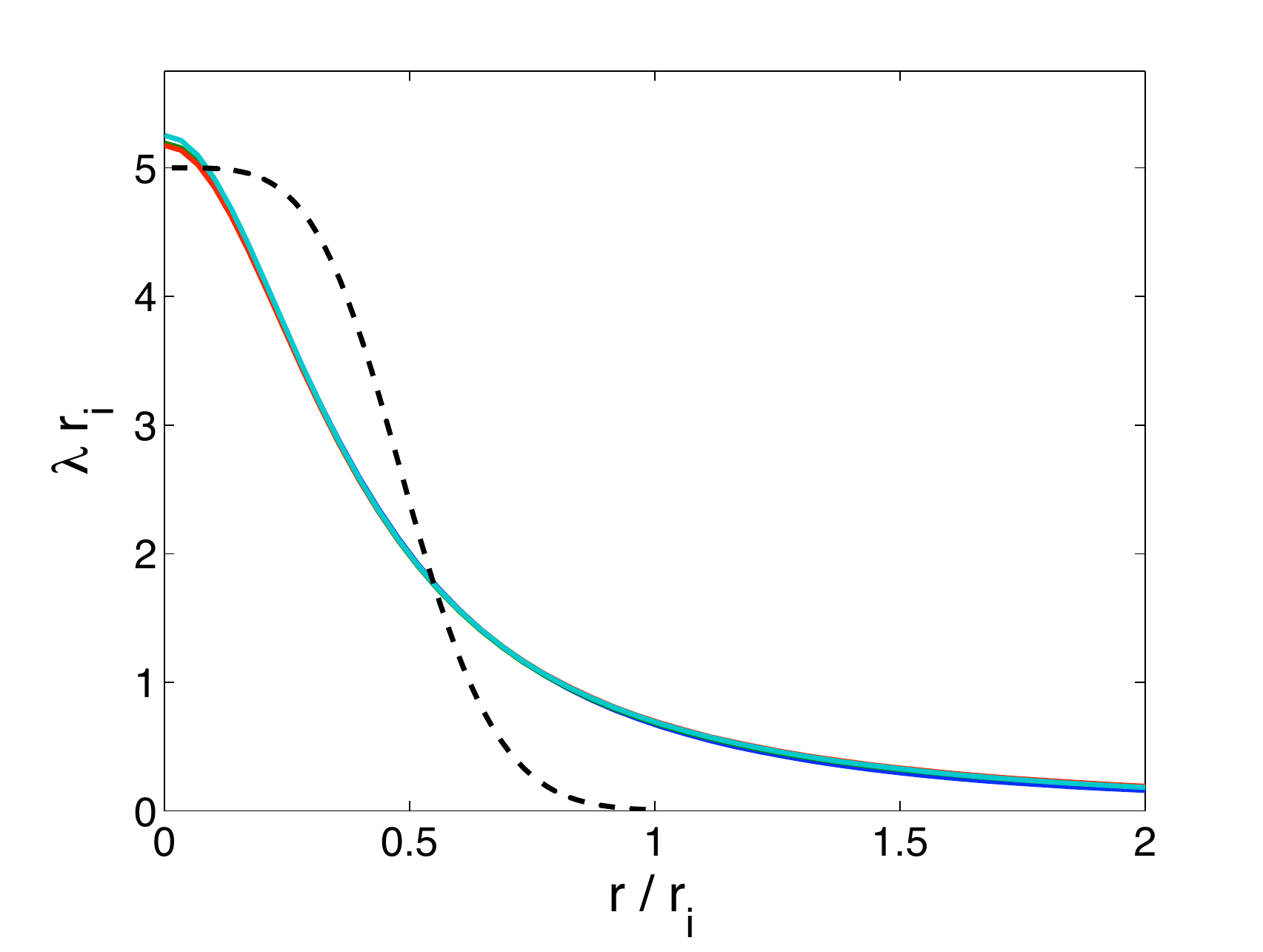}
\caption{Radial $\lambda$ profiles for $z = 20.25$, $30.14$, $40.41$, and $50.23 \; r_i$, from the nonlinear jet calculation with $\hat V_D = 4.0$, at $t = 121.7 \; T_i$ are shown as solid colored lines. The curves overlap since there is not a significant change in $\lambda$ in the $z$-direction. The $\lambda$ profile for the paramagnetic pinch with $\lambda_o = 5.0$ is shown as a dashed line.}
\label{jet_lambda_slices}
\end{figure}

A plot of radial profiles of $\lambda$ from the nonlinear jet calculation with $\hat V_D = 4.0$ at $t = 121.7 \; T_i$ is shown in Fig. \ref{jet_lambda_slices} for $z = 20.25$, $30.14$, $40.41$, and $50.23 \; r_i$. The curves overlap since there is not a significant gradient in $\lambda$ in the $z$-direction. Thus, a one dimensional equilibrium for the linear calculations is a good approximation of the $\lambda$ profiles in the nonlinear jet calculations. For comparison, the $\lambda$ profile for the paramagnetic pinch with $\lambda_o = 5.0$ is also plotted in Fig. \ref{jet_lambda_slices}. 

\begin{figure}[t]
\plotone{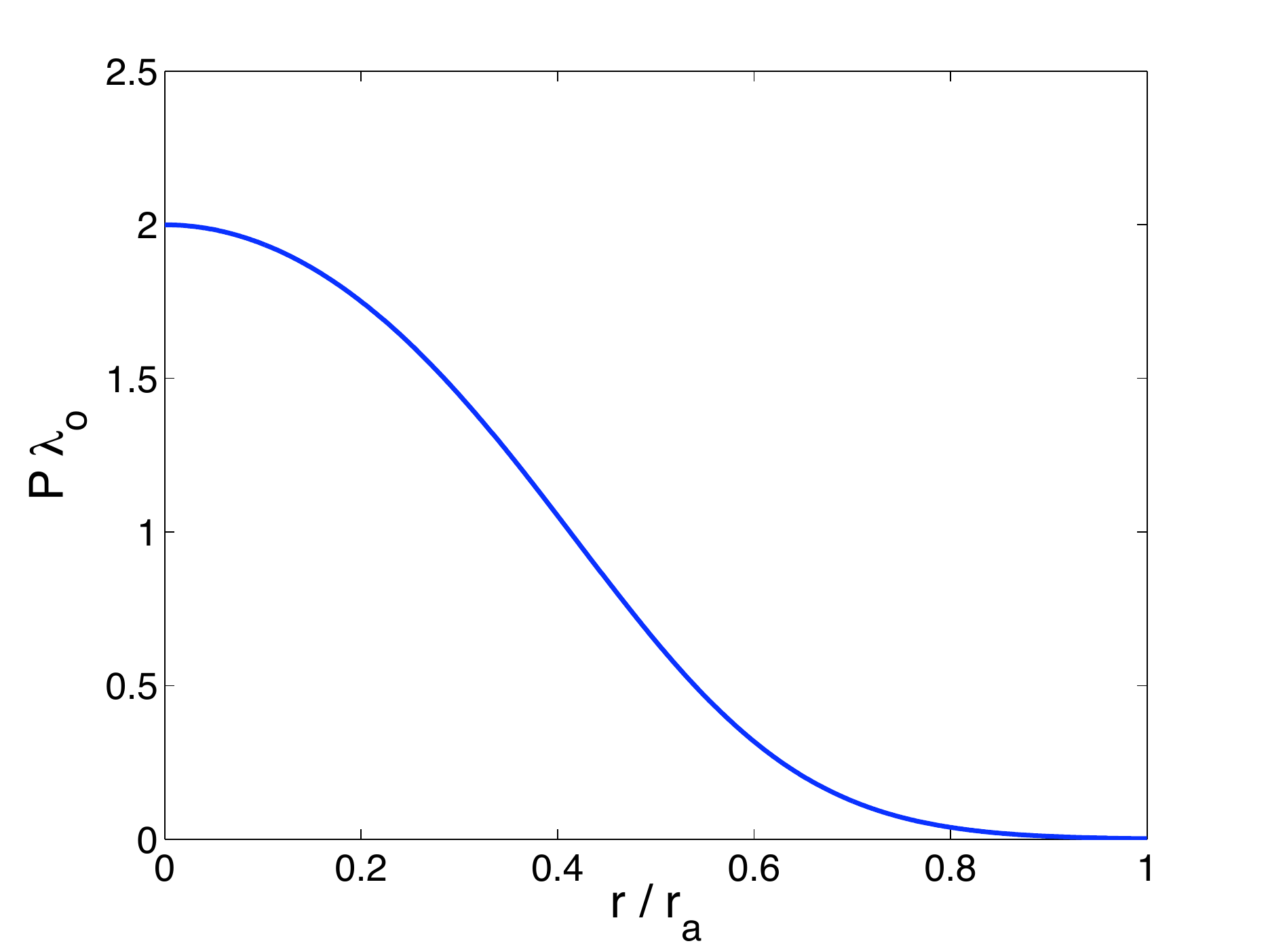}
\caption{Magnetic pitch, $P(r) = r \; B_z(r) \; B_\theta(r)^{-1}$, for the paramagnetic pinch equilibrium with $\lambda_o = 5.0$ and $\beta = 1.0$.}
\label{safety_factor}
\end{figure}

The stability of diffuse pinches, such as the paramagnetic pinch, without equilibrium flow relative to the ideal kink mode has been well studied and is known to depend on the pitch of the magnetic field, $P(r) = r \; B_z(r) \; B_\theta(r)^{-1}$. Considering eigenfunctions of the form $e^{i m \theta - i k z}$, energy analysis shows that for $m=1$ and $\frac{dp_0}{dr} = 0$, the plasma is stable if $k P(r) > 1$ or $k P(r) < (k^2 r^2 - 1) \; (3 + k^2 r^2)^{-1}$ for $r \geq 0$ and all values of $k$ \citep{Robinson:1971p4260}. When there is a region in the plasma where $k' P(r) \leq 1$ and $k' P(r) \geq (k'^2 r^2 - 1) \; (3 + k'^2 r^2)^{-1}$, there is a source of free energy for the $m=1$, $k = k'$ kink mode, and it may be unstable. When $k' P(r) < 1$ in the entire plasma, the mode is non-resonant. If there is a radius, $r_s$, in the plasma where $k' P(r_s) = 1$, $r_s$ divides the plasma into two regions; one where there is free energy for the kink, and one where there is not; and the mode is called resonant. For the paramagnetic pinch, $P(r)$ decreases monotonically, and there is free energy for the kink in the region with $r > r_s$. Since the free energy for the kink is at radii larger than $r_s$, the stabilizing effect of the conducting boundary at $r = r_a$ affects both resonant and non-resonant modes. The magnetic pitch profile of the equilibrium used here is shown in Fig. \ref{safety_factor}. Since $P(r = 0) = 2.0 \; \lambda_o^{-1}$, $k \geq 0.5 \: \lambda_o$ modes are resonant and $k < 0.5 \: \lambda_o$ modes are non-resonant. 

\begin{figure}[t]
\plotone{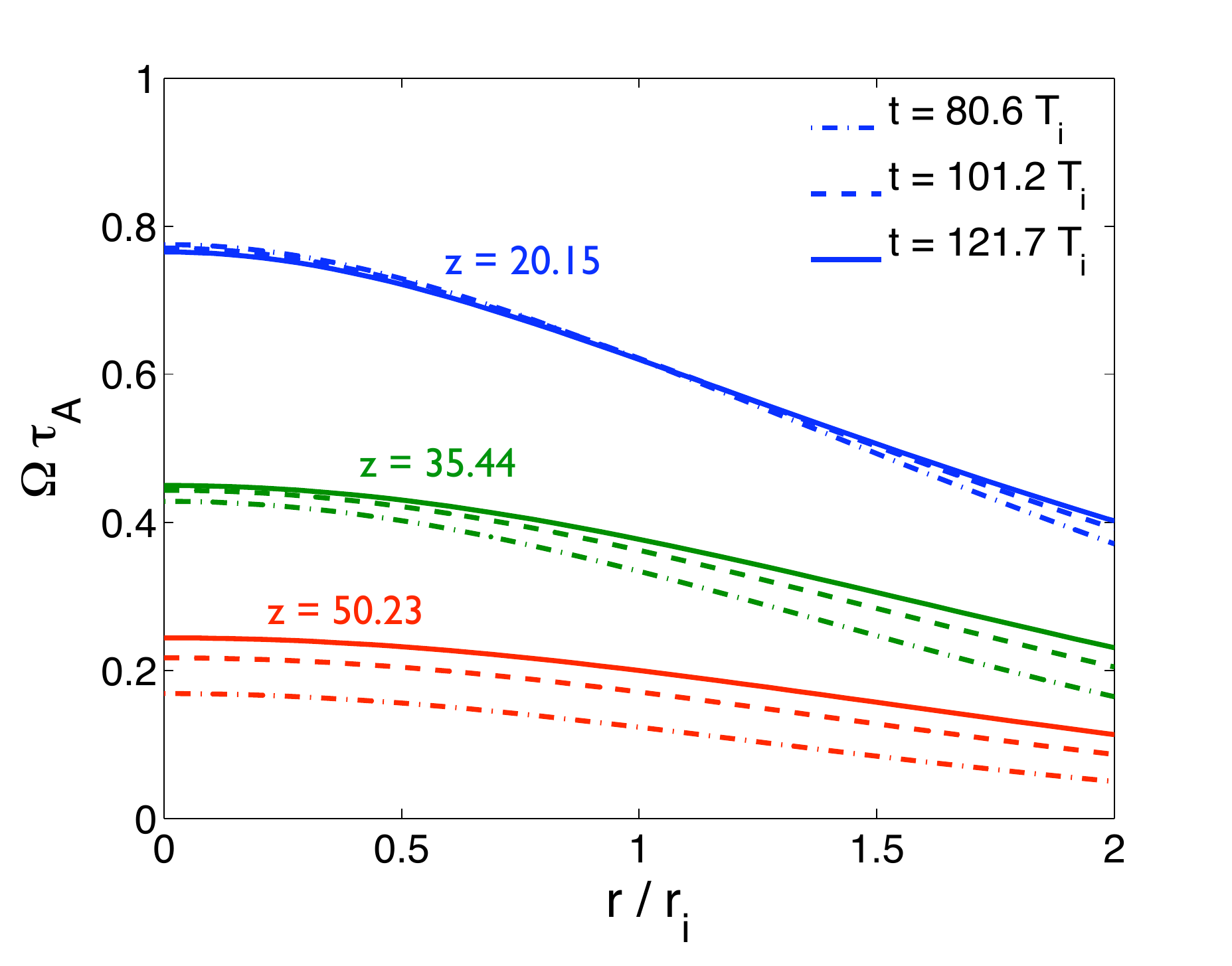}
\caption{Radial profiles of the jet rotation frequency for $z = 20.25$, $33.44$, and $50.23 \; r_i$, from the nonlinear jet calculation with $\hat V_D = 4.0$, at times $t = 80.6$, $101.2$, $121.7 \; T_i$. The rotation frequency, $\Omega$, is given by $\Omega = v_\theta \; r^{-1}$.}
\label{jet_rot_freq_slices}
\end{figure}

To examine the effect of jet rotation, we consider MHD equilibria with rigid rotation in the $\theta$-direction, and use $\Omega$ to denote the rotation frequency. While previous studies have shown that sheared flow is more efficient at stabilizing the kink mode \citep{Wanex:2005p1603}, our nonlinear computations show little azimuthal shear in the vicinity of the the jet. Radial profiles of the jet rotation frequency from the nonlinear jet simulation with $\hat V_D = 4.0$ at various times and axial positions are shown in Fig. \ref{jet_rot_freq_slices}. As time increases, the jet rotation frequency reaches a steady state at higher axial positions along the length of the column. For all values of $z$, the rotation frequency is uniform to within $20 \%$ across the radius of the jet, which has a width of $r \leq 1$, and as the column propagates, the rotation frequency flattens. Thus, rigid rotation is a reasonable simplification. 

The paramagnetic pinch is often considered to be a force-free equilibrium in which the current is purely parallel to the magnetic field. However, equilibrium azimuthal flow breaks the force-free nature, since the centrifugal force of the flow must be balanced by another MHD force. Two choices of force balance are considered here. The first, labeled `magnetic-balance', balances the centrifugal force against the force from the perpendicular current,

\begin{equation}
\mathbf J_0 \bm \times \mathbf B_0 = - \rho_0 \Omega^2 \mathbf r,
\label{stabalization_3}
\end{equation}

\noindent
and the parallel current is unchanged. Thus, while the current profile is modified by the introduction of the rotation, the $\lambda(r)$ profile, which is related to the free-energy source for the kink mode, is unaffected. The second force-balance model, labeled `pressure-balance', balances the centrifugal force against the equilibrium pressure,

\begin{equation}
\bm \nabla p_0 = \rho_0 \Omega^2 \mathbf r.
\label{stabalization_4}
\end{equation}

\noindent
For this case the current profile is unchanged by the introduction of the rotation. However, as the rotation increases, the pressure profile becomes increasingly hollow in the sense that it peaks on the edge of the plasma, which can have a stabilizing effect \citep{Freidberg:1987p2544pg259}. The equilibrium pressure is characterized by the plasma $\beta$ on the central axis. The choice of the values for $\lambda_o$ and $\beta$ is motivated by our nonlinear jet calculations, giving $\lambda_o = 5.0$ and $\beta = 1.0$. 

\begin{figure}[t]
\plotone{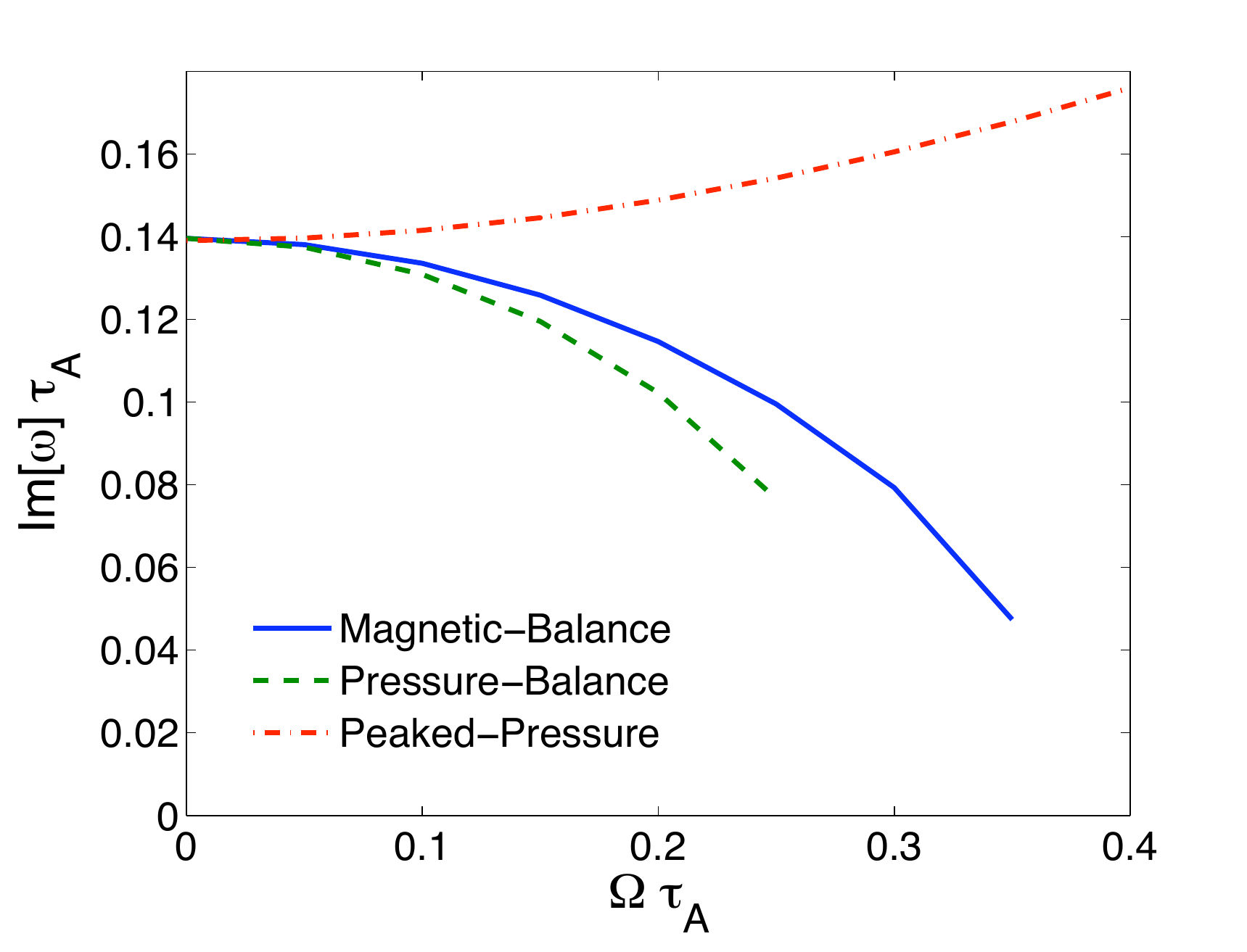}
\caption{Growth rates from linear initial value calculations for the magnetic-balance and pressure-balance models, and with an equilibrium pressure profile which replaces the centrifugal force (peaked-pressure model). For the peaked-pressure model, there is no equilibrium rotation; instead, the rotationally equivalent pressure is given by $p_0(\Omega,r) = \beta - \Omega^2 r^2 / 2$.}
\label{nim_lin_gr}
\end{figure}

Our numerically computed growth rate of the $m=1$, $k = 0.4 \; \lambda_o$ kink mode as a function of equilibrium rotation frequency, for both force balance models, is plotted in Fig. \ref{nim_lin_gr}. The results show that the growth rate of the mode decreases as rotation increases for both force balance models. The growth rate decreases somewhat faster in the pressure-balance model than in the magnetic-balance model, which we surmise is a result of the additional stabilizing effect of the hollow pressure profile in the pressure-balance model. While the results point to rotation as the important stabilizing mechanism, force-balance requires changes to the pressure profile or the perpendicular current profile as rotation is increased. To examine the effect of modifying the equilibrium forces to balance the centrifugal force from the rotation, a plasma which has the same equilibrium current as the magnetic-balance model, but without rotation, is considered. Here, the equilibrium pressure gradient replaces the centrifugal force by defining a profile which is peaked on the central axis. The resulting growth rate is also plotted in Fig. \ref{nim_lin_gr}. As the pressure gradient increases, the growth rate of the kink mode increases. This result confirms that rotation is the stabilizing influence in the $\Omega$-scans.

\begin{figure}[t]
\plotone{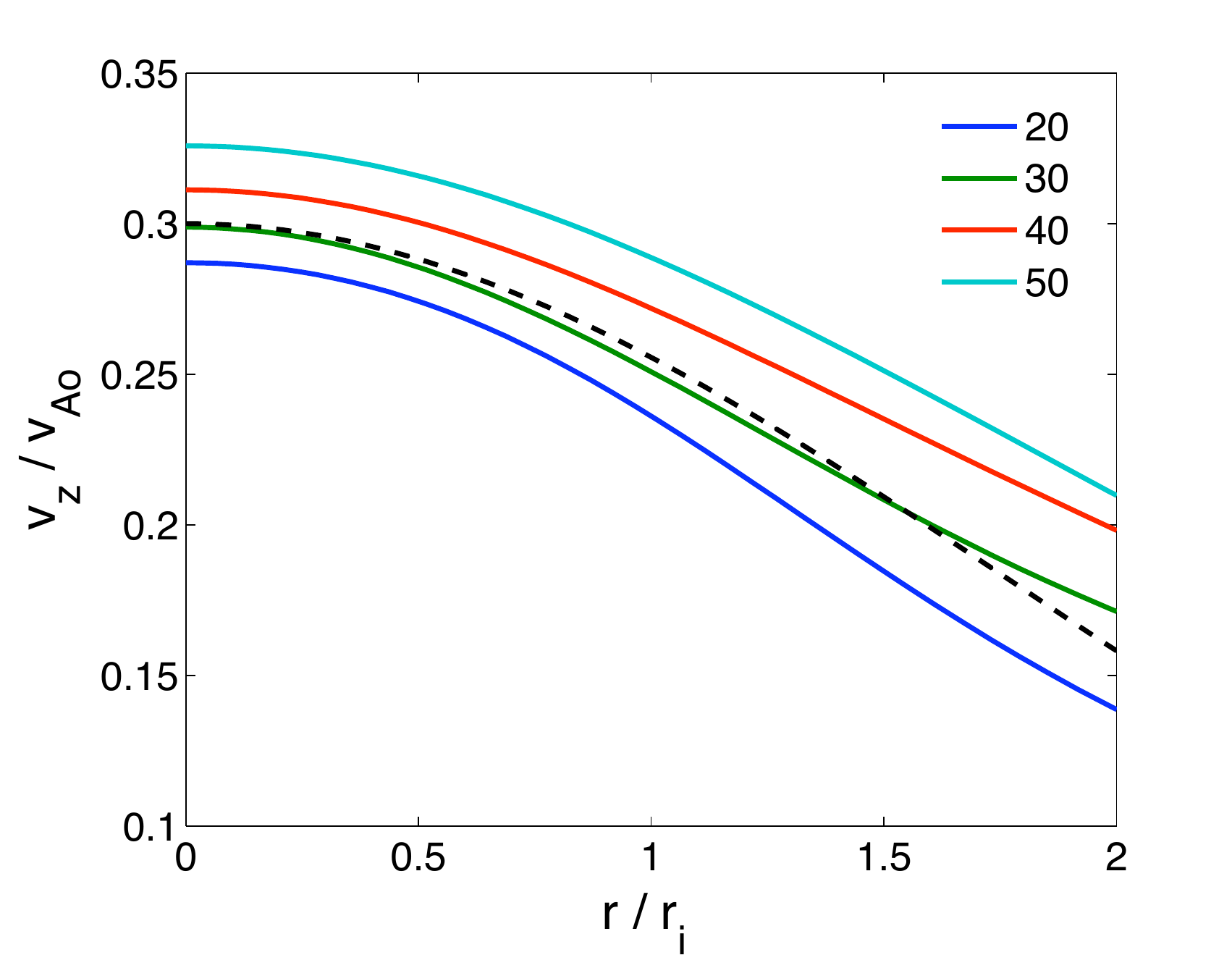}
\caption{Axial flow profiles from the stable nonlinear jet simulation with $\hat V_D = 4.0$ at $t = 121.7 \; T_i$ at the indicated axial locations. The black dashed line shows the Gaussian flow profile used in the linear calculations with $v_M = 0.3 \; v_A$ and $w_g = 25.0 \; \lambda_o^{-1}$.}
\label{jet_vz}
\end{figure}

\begin{figure}[t]
\plotone{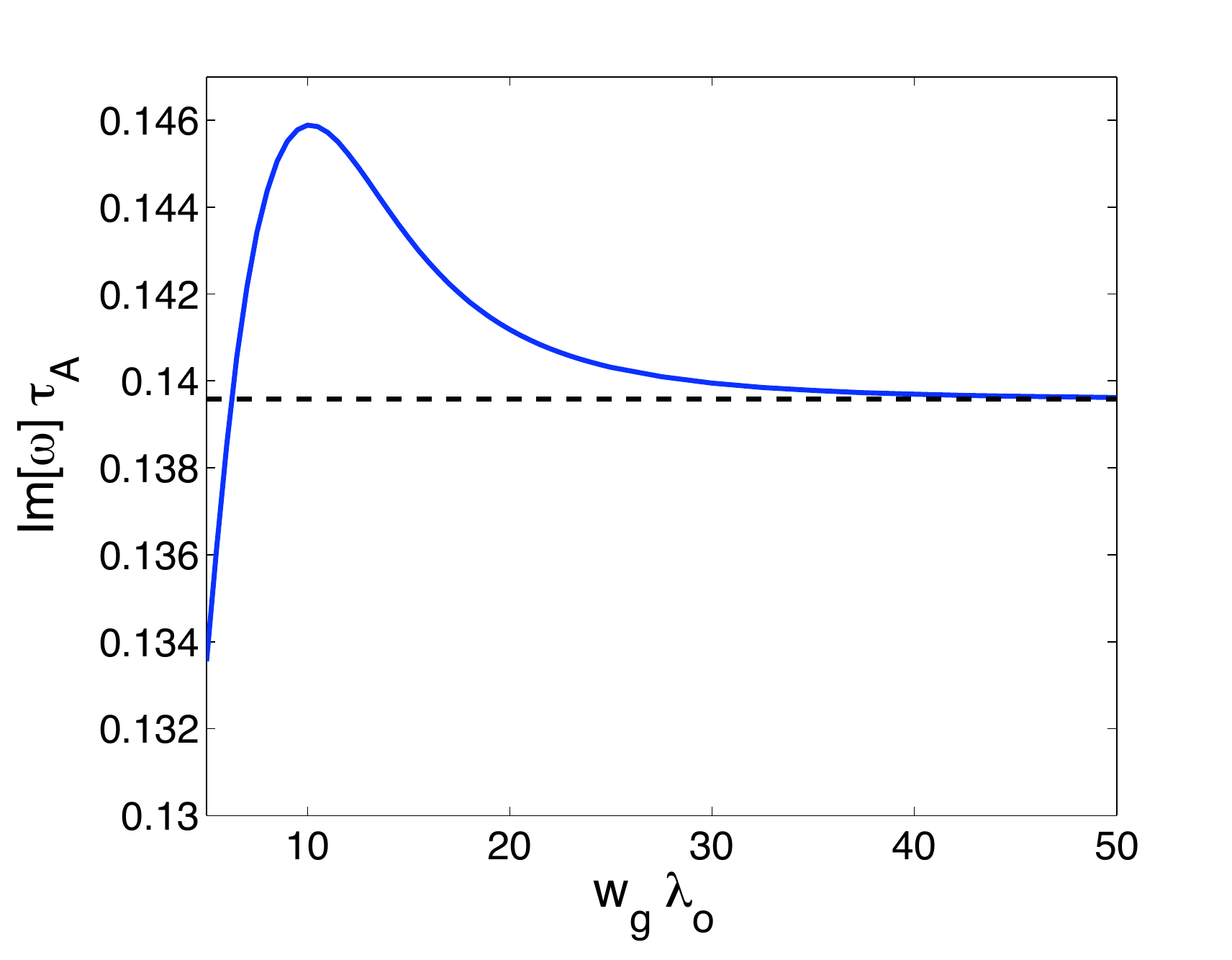}
\caption{Growth rates from linear initial value calculations with equilibrium axial flow as a function of the Gaussian $v_z$ profile width. The black dashed line shows the growth rate without equilibrium flow.}
\label{gamma_vs_wg}
\end{figure}

Previous theoretical and experimental studies show that sheared axial flow can stabilize the kink mode in a cylindrical plasma \citep{Shumlak:1995p2833, Shumlak:2003p1156}. Thus, we consider what effect axial flow has on jet stability in the nonlinear simulations via linear initial value calculations with equilibrium axial flow. Non-rotating force-free paramagnetic pinch equilibria with Gaussian axial flow profiles, given by $v_z(r) = v_M \: e^{-\left(2 r / w_g \right)^2}$, are considered. Motivated by the axial flow profiles in the nonlinear jet simulations, we choose $v_M = 0.3 \; v_A$ and consider a range of $w_g$ from $5.0$ to $50.0 \; \lambda_o^{-1}$, where smaller values of $w_g$ correspond to larger flow shear. Axial flow profiles from the stable $\hat V_D = 4.0$ jet simulation and the Gaussian profile used for the linear calculation with $w_g = 25.0 \; \lambda_o^{-1}$ are shown in Fig. \ref{jet_vz}. Growth rates of the kink mode as a function of $w_g$ are plotted in Fig. \ref{gamma_vs_wg}. A flow shear range comparable to that considered by \citet{Shumlak:1995p2833} is considered, but the change in the kink growth rate is less than $6.0 \%$. We attribute this to the difference in the equilibria considered here and that examined by \citet{Shumlak:1995p2833}. Based on these results, we conclude that axial flow does not significantly influence the stability of the magnetic column in our nonlinear jet simulations.

\section{Linear Eigenvalue Calculations}
\label{levc}

The results of the linear initial value calculations indicate that the nonlinear simulations remain robust to the kink mode for high rotation rates of the accretion disk because of the rotation of the jet itself. To further examine the effect of azimuthal rotation on the kink mode, we investigate the linear ideal MHD spectrum for rotating paramagnetic equilibria. This eigenmode analysis helps us develop physical insight into the effect of rotation, which is difficult to obtain from the initial value calculations. The theory considers a cylindrical domain which is periodic in the $z$-direction with the one-dimensional rigid-rotation equilibria described in Section \ref{livc}.

\subsection{Linear Eigenvalue Theory}
The simplest approach in considering an MHD equilibrium with flow is to work in a Lagrangian representation. Assuming perturbations to the equilibrium depend on time as $e^{-i \omega t}$, the linearized MHD equation of motion is given by

\begin{eqnarray}
& -\rho_0  \omega^2 \bm \xi - 2 i \rho_0 \omega \mathbf v_0 \bm \cdot \bm \nabla \bm \xi \nonumber \\
& + \rho_0 \mathbf v_0 \bm \cdot \bm \nabla (\mathbf v_0 \bm \cdot \bm \nabla \bm \xi) = \mathbf F(\bm \xi),
\label{lin_th_lag_frame_1}
\end{eqnarray}

\noindent
where from this point, fields without subscripts represent perturbations and are assumed to be small \citep{Freiman:1960p1471, Waelbroeck:1996p1472}. The plasma displacement, $\bm \xi$, is defined by

\begin{equation}
\mathbf v = \frac{\partial \bm \xi}{\partial t}.
\end{equation}

\noindent
The linear force operator, $\mathbf F( \bm \xi)$, is given by

\begin{eqnarray}
\mathbf F( \bm \xi )= -\bm \nabla p + \frac{1}{\mu_o}\mathbf J_0 \bm \times \mathbf B + \frac{1}{\mu_o}(\bm \nabla \bm \times \mathbf B) \bm \times \mathbf B_0 \nonumber \\
 + \bm \nabla \bm \cdot (\rho_0 \bm \xi \; \mathbf v_0 \bm \cdot \bm \nabla \mathbf v_0),
\label{lin_th_lag_frame_2}
\end{eqnarray}

\noindent
where the perturbed magnetic field and the perturbed pressure are given by

\begin{equation}
\mathbf B = \bm \nabla \bm \times (\bm \xi \bm \times \mathbf B_0),
\label{lin_th_lag_frame_3}
\end{equation}

\begin{equation}
p = -(\bm \xi \bm \cdot \bm \nabla p_0 + \gamma p_0 \bm \nabla \bm \cdot \bm \xi).
\label{lin_th_lag_frame_4}
\end{equation}

As a consistency check, we also evaluate the spectra derived from an Eulerian frame of reference by including equilibrium flow in the definition of the Lagrangian displacement vector, $\bm \xi$, which satisfies \citep{Chandrasekhar:1961p3747}

\begin{equation}
\mathbf v = \frac{\partial \bm \xi}{\partial t} + \bm \nabla \bm \times (\bm \xi \bm \times \mathbf v_0).
\label{lin_eu_frame_1}
\end{equation}

\noindent
Linearizing the MHD equations in the Eulerian frame with rigid equilibrium rotation gives the following momentum equation, force operator, induction equation, and pressure equation respectively,

\begin{eqnarray}
- \omega_D^2 \bm \xi - 2 i \Omega \omega_D (\bm{\hat z} \bm \times \bm \xi) +  r \Omega^2 (\bm \nabla \bm \cdot \bm \xi) \nonumber \\ 
\left[ \left( 3 + \frac{m \Omega}{\omega_D} \right) \bm{\hat r} + i \frac{\omega_D}{\Omega} \bm{\hat \theta} \right] = \frac{1}{\rho_0} \mathbf F( \bm \xi ) ,
 \label{lin_eu_frame_2.1}
\end{eqnarray}

\begin{eqnarray}
\mathbf F( \bm \xi ) = & \frac{1}{\mu_o} \left( \mathbf B_0 \bm \cdot \bm \nabla \mathbf B + \mathbf B \bm \cdot \bm \nabla \mathbf B_0 \right) - \nonumber \\ 
& \bm \nabla \left( p + \frac{1}{\mu_o} \mathbf B \bm \cdot \mathbf B_0 \right),
 \label{lin_eu_frame_2.2}
\end{eqnarray}

\begin{equation}
\mathbf B = \bm \nabla \bm \times (\bm \xi \bm \times \mathbf B_0) + \frac{\Omega {B_0}_z}{\omega_D} (\bm \nabla \bm \cdot \bm \xi) (r k \bm{\hat \theta} - m \bm{\hat z}),
\end{equation}

\begin{equation}
p = - (\bm \xi \bm \cdot \bm{\hat r}) \frac{dp_0}{dr} - \gamma p_0 \left( 1 + \frac{m \Omega}{\omega_D} \right) (\bm \nabla \bm \cdot \bm \xi),
\label{lin_eu_frame_2}
\end{equation}

\noindent
where $\omega_D = \omega - m \Omega$ is the Doppler shifted eigenfrequency. 

Generalizing the analysis in Ref. \citep{Freidberg:1987p2544pg473}, the linearized equations in either reference frame are reduced to a pair of coupled first-order differential equations for the radial plasma displacement, $\xi_r$, and the total perturbed plasma pressure, $\tilde P = p + B \; B_0 / \mu_o$. Assuming spatial dependence of the perturbed fields of the form $e^{i m \theta - i k z}$, Eqs. \ref{lin_th_lag_frame_1}-\ref{lin_th_lag_frame_4} and Eqs. \ref{lin_eu_frame_2.1}-\ref{lin_eu_frame_2} become systems of ordinary differential equations (ODE's) with respect to the $r$-coordinate. By considering the projection of Eqs. \ref{lin_th_lag_frame_1} and \ref{lin_eu_frame_2.1} in the $\bm{\hat b}$ and $\bm{\hat \eta}$ directions, where $\bm{\hat b} = \frac{\mathbf B_0}{| \mathbf B_0 |}$ and $\bm{\hat \eta} = \bm{\hat b} \bm \times \bm{\hat r}$, the $\bm{\hat b}$ and $\bm{\hat \eta}$ components of the plasma displacement can be solved analytically. Substituting these results into Eqs. \ref{lin_th_lag_frame_1} and \ref{lin_th_lag_frame_4} and Eqs. \ref{lin_eu_frame_2.1} and \ref{lin_eu_frame_2} produces sets of coupled ODE's with the same general form,

\begin{equation}
\underline{\underline{A}}(r,\omega) \; \frac{d}{dr} \left(\begin{array}{c}r \xi_r \\ \tilde P \end{array}\right) = \underline{\underline{B}}(r,\omega)  \;  \left(\begin{array}{c}r \xi_r \\ \tilde P \end{array}\right),
\label{lin_th_lag_frame_5}
\end{equation}

\noindent
in both reference frames.

We consider the plasma to be surrounded by a conducting shell at the radius $r = r_a$ by defining $\xi_r(r_a) = 0$. The regularity condition at $r = 0$ is imposed by the cylindrical geometry of the domain. Expansion of $\xi_r$ in a power series for small values of $r$ shows that regular solutions satisfy $\xi_r \propto r^{m-1}$. Equation \ref{lin_th_lag_frame_5} coupled with these boundary conditions defines an eigenvalue problem with $\omega$ as the eigenvalue.

It should be noted that while the form of this eigenvalue equation is the same in both reference frames, the ODE coefficient matrices $\underline{\underline{A}}$ and $\underline{\underline{B}}$ are unique to each frame. Equation \ref{lin_th_lag_frame_5} is derived for a general equilibrium flow in a Lagrangian frame, and the coefficients can be found in \citet{Bondeson:1987p2834}. The ODE coefficients for a plasma equilibrium with rigid rotation and uniform axial flow in an Eulerian frame can be found in \citet{Appl:1992p2913}.

Due to the complexity of the ODE coefficients in Eq. \ref{lin_th_lag_frame_5}, we use a shooting method to solve the eigenvalue problem. A value is chosen for $\omega$, and Eq. \ref{lin_th_lag_frame_5} is numerically integrated from $r = 0$ to $r = r_a$ using fourth-order Runge-Kutta integration. The choice of $\omega$ is varied until the eigenfunction satisfies $\xi_r(r_a) = 0$. A Newton-Raphson method is used to search the $\omega$-parameter space for functions that satisfy this boundary condition.

In the absence of equilibrium flow, the MHD force operators in Eqs. \ref{lin_th_lag_frame_2} and \ref{lin_eu_frame_2.2} are self-adjoint, and $\omega$ is either purely real or purely imaginary \citep{Freidberg:1987p2544pg242}. With the introduction of equilibrium flow, the force operator is no longer self-adjoint, and $\omega$ and $\bm \xi(r)$ can be complex \citep{Freiman:1960p1471}. The real component of the eigenvalue, $\Re[\omega]$, gives the oscillation frequency of the eigenmode, and the imaginary component, $\Im[\omega]$, determines its growth or decay rate. The Newton-Raphson method employed here is generalized to search the complex parameter space \citep{Press:2007p3009}. While Newton-Raphson readily generalizes to multiple dimensions, it converges only if the initial guess for the root is in the vicinity of the actual root. Since $\omega$ is either purely real or purely imaginary without the equilibrium flow, Newton-Raphson is used in a one-dimensional space to find $\Im[\omega]$ with $\Omega = 0$ for a given mode. The $\Omega = 0$ result is then used as an initial guess for a nearby equilibrium with flow, and that result is used as an initial guess for a slightly larger value of $\Omega$. This process is repeated for increasing values of $\Omega$.

\subsection{Linear Eigenvalue Results}
\label{linear_eigenvalue_results}

\begin{figure}[t]
\plotone{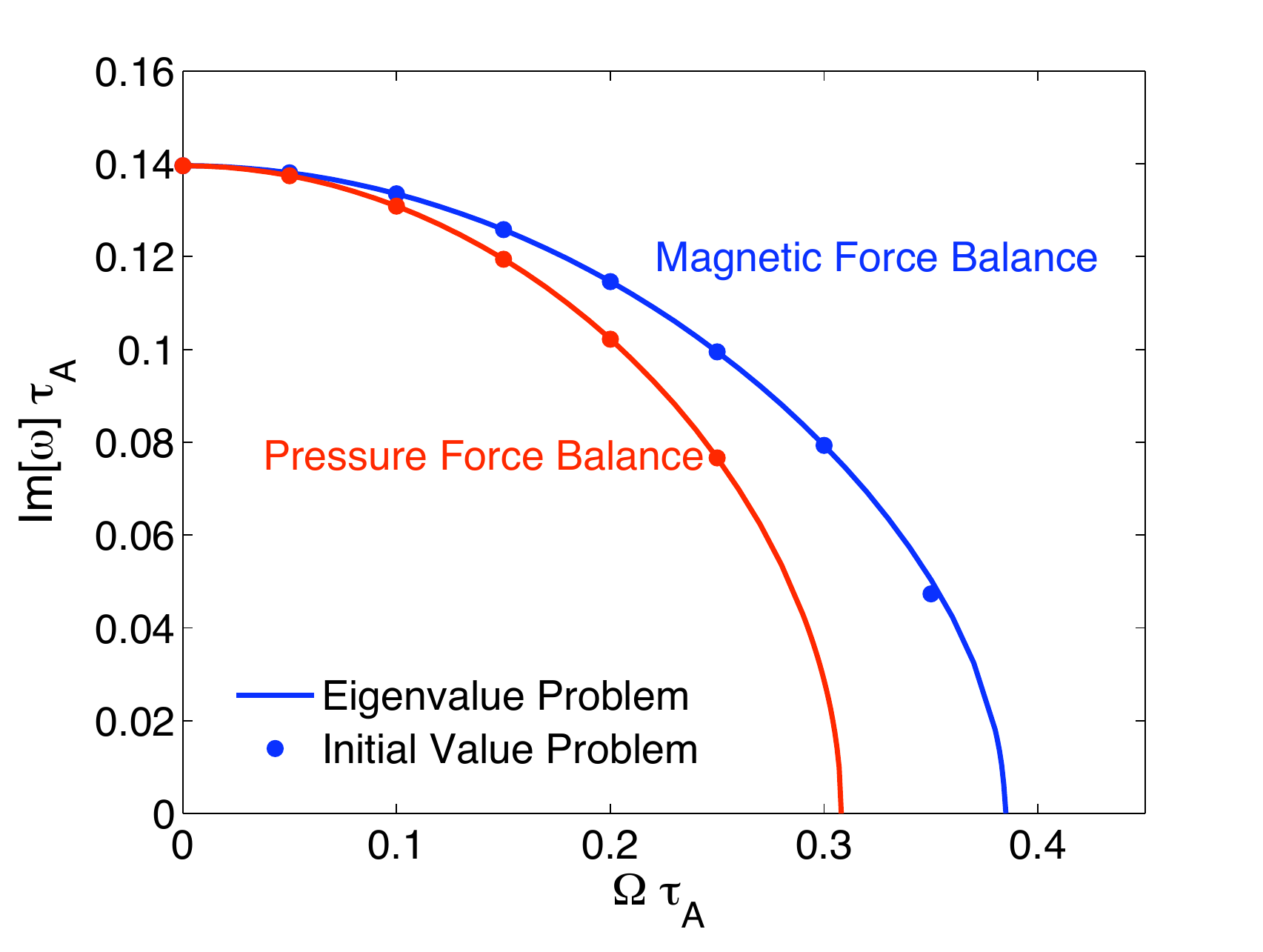}
\caption{Growth rates of the non-resonant $m=1$, $k=0.4 \; \lambda_o$ kink mode as a function of the equilibrium rotation frequency, $\Omega$, from Lagrangian and Eulerian eigenvalue calculations, and from the linear initial value calculations. }
\label{gamma_vs_rotation}
\end{figure}

\begin{figure}[t]
\plotone{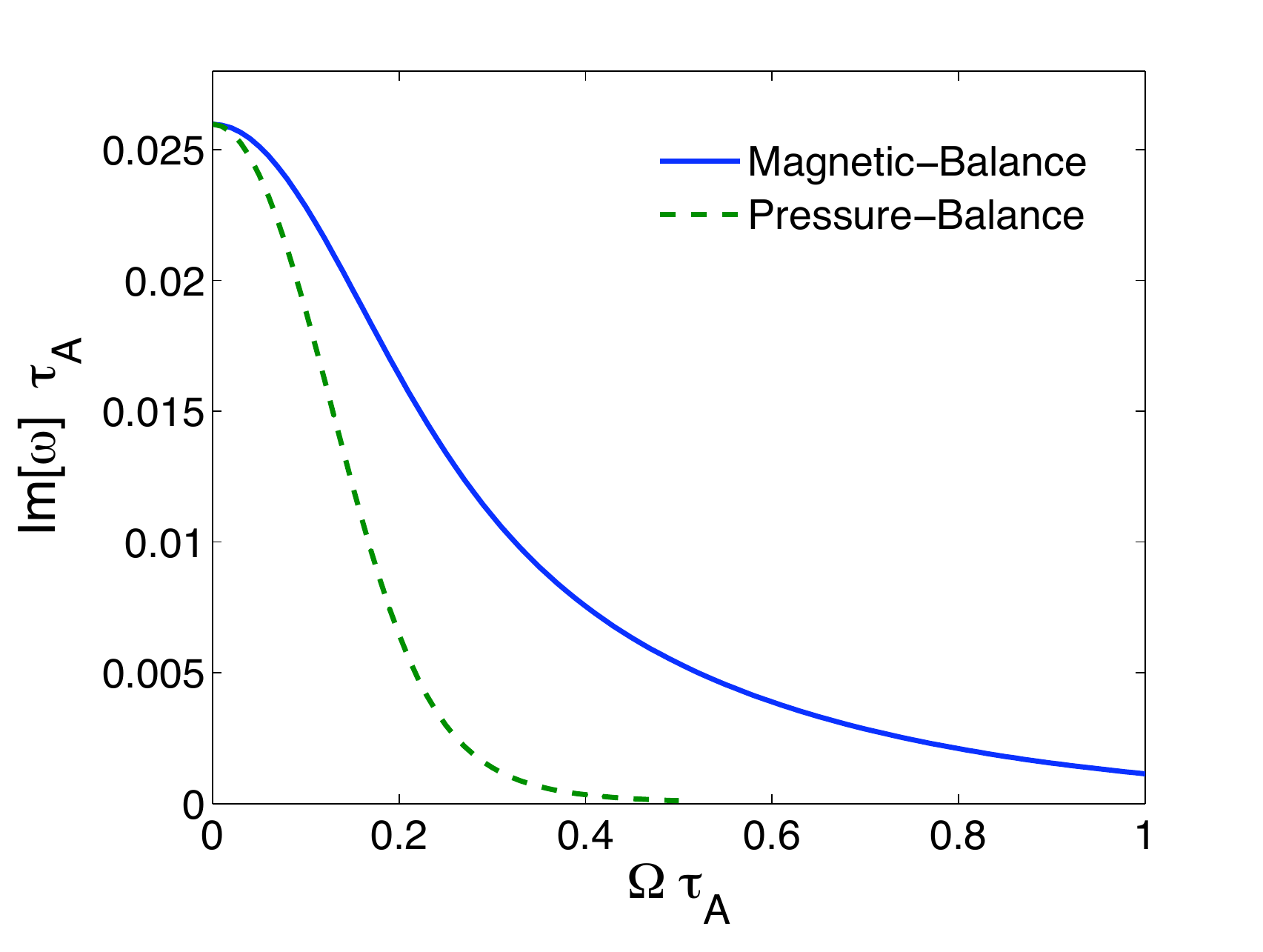}
\caption{Growth rates of the resonant $m=1$, $k=0.6 \; \lambda_o$ kink mode as a function of the equilibrium rotation frequency, $\Omega$. }
\label{resonant_growth_rates}
\end{figure}

Results of the eigenmode analysis and growth rates from the initial value formulation of Sec. \ref{livc} for the non-resonant $m=1$, $k = 0.4 \; \lambda_o$ kink mode can be seen in Fig. \ref{gamma_vs_rotation}. Here, calculations are shown for both both force-balance models in both reference frames. The curves from the two reference frames are indistinguishable in this plot, and comparison of the eigenvalue formulation and the initial value formulation of the problem are shown to be in agreement. These results show that as $\Omega$ increases, the growth rate of the kink mode decreases and is stable with sufficient rotation. We note that the marginal rotation period is larger than the Alfv\'en propagation time, i.e. Alfv\'enic flow within the cylinder is not required for stabilization.

We also examine the effect of rotation on resonant kink modes via the eigenvalue formulation. Growth rates for the $m=1$, $k=0.6 \: \lambda_o$ mode can be seen in Fig. \ref{resonant_growth_rates}. While equilibrium rotation fully stabilizes the non-resonant kink mode described previously, rotation only reduces the growth rate of the resonant mode and does not completely stabilize it.

\begin{figure}[t]
\plotone{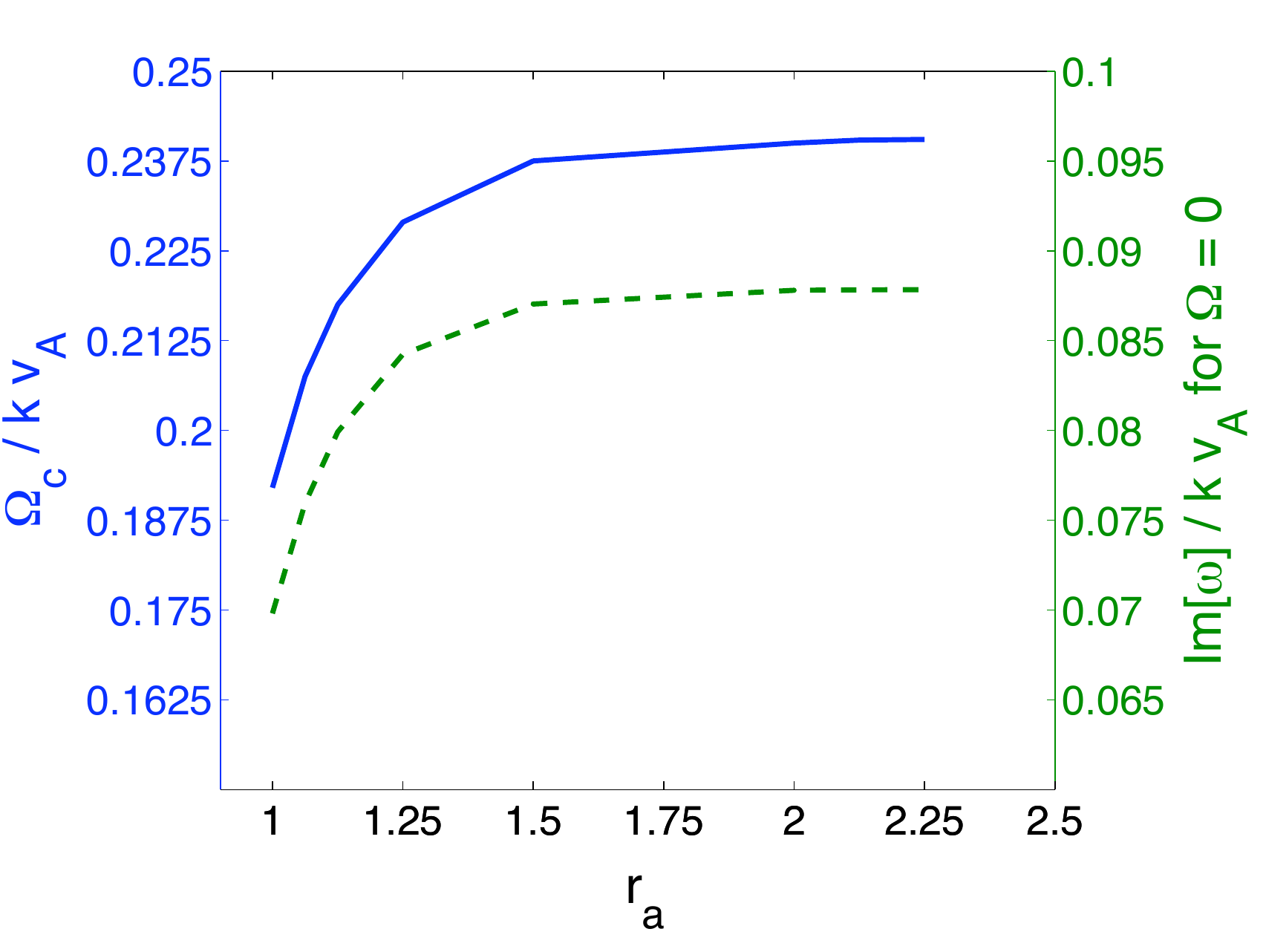}
\caption{Critical rotation frequency for stabilization of the kink mode (Solid Line), and growth rate of the kink mode for $\Omega = 0$ (Dotted Line), as a function of the outer radial boundary, $r_a$, for the $m=1$, $k = 0.4 \; \lambda_o$ kink mode.}
\label{crit_rot_vs_rmax}
\end{figure}

The eigenmode solutions treat the radial boundary at $r = r_a$ as a solid wall by setting $\xi(r_a) = 0$. However, there is no close boundary surrounding the plasma column in the nonlinear jet simulations. To evaluate the influence of the wall location, we recompute the eigenvalues as $r_a$ is varied. The critical rotation frequency, $\Omega_c$, for stabilization of the $m=1$, $k = 0.4 \; \lambda_o$ kink mode as a function of $r_a$ is plotted in Fig. \ref{crit_rot_vs_rmax}. The resulting critical rotation frequency asymptotically approaches  the value $\Omega_c = 0.24 \; (k \; v_A)^{-1}$, indicating that the stabilizing effect of the rotation remains as $r_a \rightarrow \infty$. The growth rate with $\Omega = 0$, $\gamma_o(r_a)$, as a function of $r_a$ is also plotted in Fig. \ref{crit_rot_vs_rmax}. The $\Omega_c(r_a)$ and $\gamma_o(r_a)$ curves follow the same asymptotic trend, implying that the dependence of $\Omega_c$ on $r_a$ is  related to the free energy of the kink mode and not due to any changes in the stabilizing influence of rotation. 

\begin{figure}[t]
\plotone{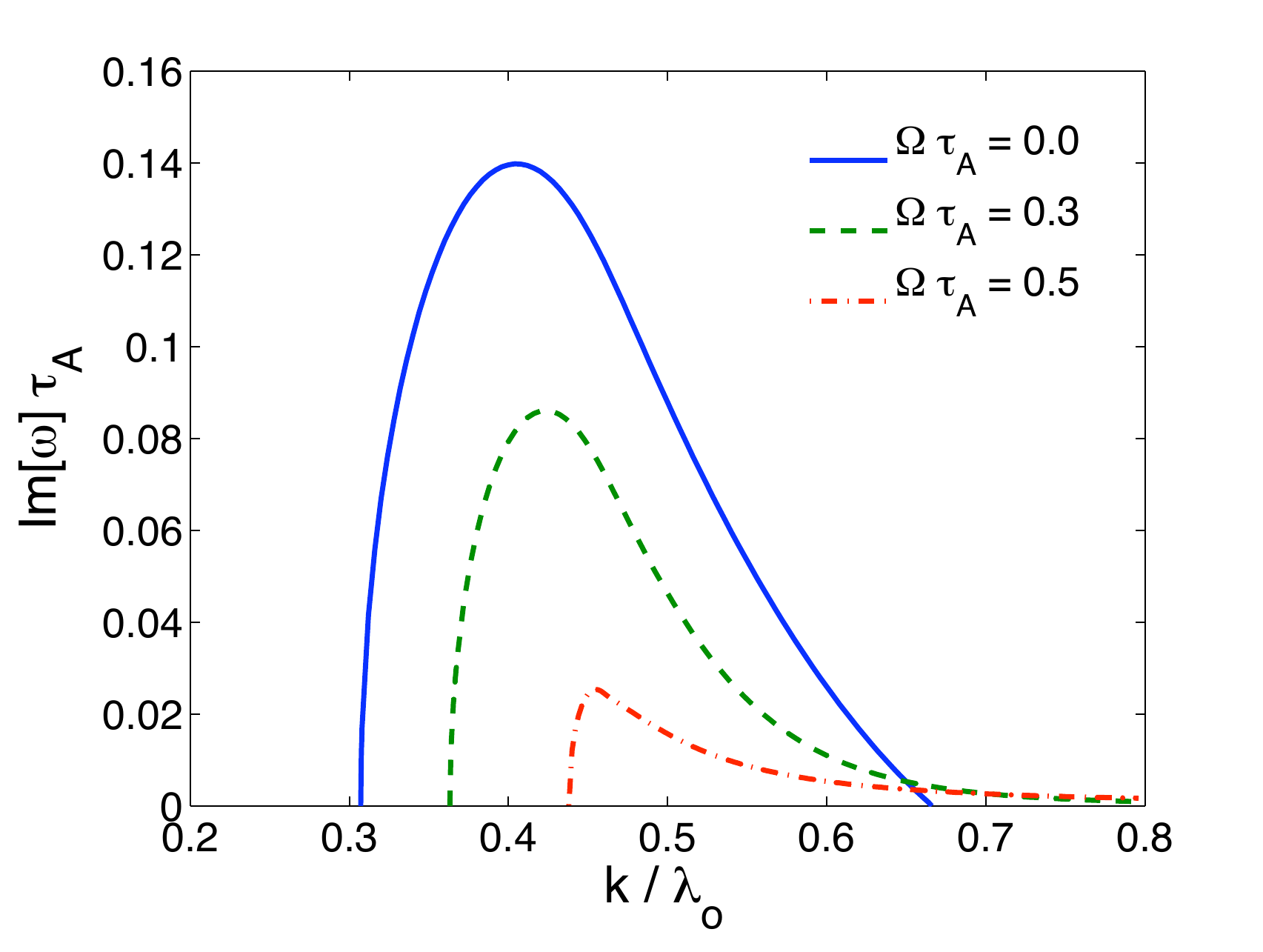}
\caption{Growth rates of the $m=1$ kink mode as a function of the axial wave number, $k$, for various equilibrium rotation frequencies.}
\label{gr_vs_k}
\end{figure}

The linear eigenvalue formulation allows for the examination of a range of axial wave numbers. The growth rate of the $m=1$ kink mode as a function of $k$ for various values of $\Omega$ is calculated, and the results are shown in Fig. \ref{gr_vs_k}. Without equilibrium rotation, there are lower and upper bounds on the unstable values of $k$. Both the upper and the lower bound increase with increasing rotation. For the equilibria considered here, modes with $k < 0.5 \: \lambda_o$ are non-resonant, and modes with $k \geq 0.5 \: \lambda_o$ are resonant. While the range in $k$-space of unstable non-resonant kink modes decreases with increasing rotation, the range of unstable resonant modes broadens with small growth rates on the order of $10^{-3} \; \tau_A^{-1}$. 

We have explored a range of $\beta$ values to examine the effect of equilibrium thermal pressure on rotational stabilization. For moderate values of $\beta$, rotational stabilization is observed to be independent of $\beta$. However, for low $\beta$-values ($\beta \leq 0.06$) rigid rotation destabilizes the kink mode for $\Omega \gtrsim 0.4 \; \tau_A^{-1}$. The destabilized modes are compressible with a $\theta$-component of $\vec \xi$ that is much larger than the other components. Thus, these modes are stabilized by equilibrium pressure for the moderate values of $\beta$ relevant to extragalactic jet systems.

\begin{figure}[t]
\plotone{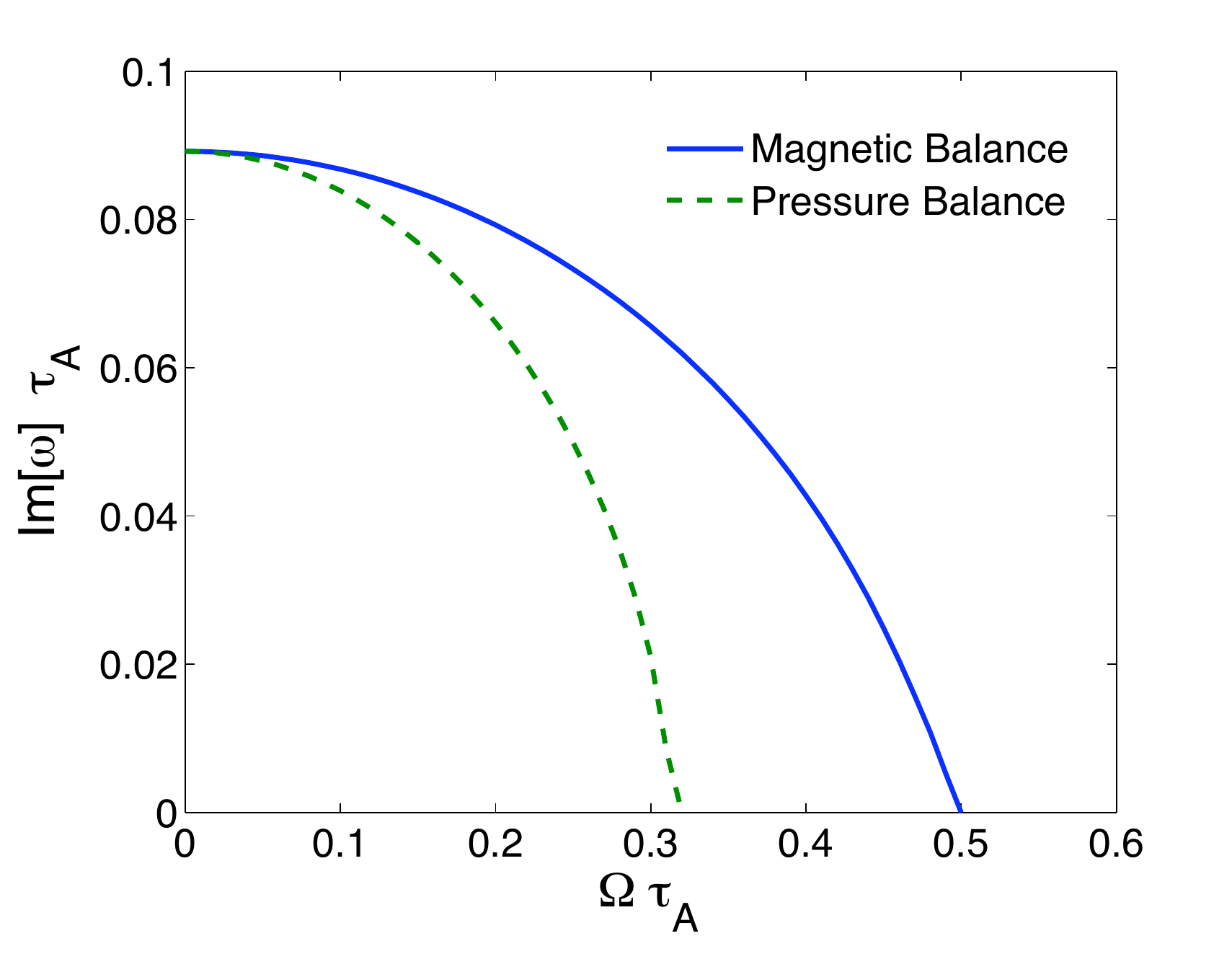}
\caption{Growth rate of the $m=1$, $k=1.0 \; r_a^{-1}$ non-resonant kink mode for monotonically increasing magnetic pitch equilibria as a function of equilibrium rotation.}
\label{q_increasing_gamma_vs_rotation}
\end{figure}

The choice of initial and disk boundary conditions in simulations of jet formation can have a profound effect on the magnetic pitch profile, $P(r)$ \citep{Moll:2008p4392}. Thus far, we have considered only equilibria with monotonically decreasing $P(r)$ as is observed in the nonlinear jet simulations discussed in Sec. \ref{nonlinear_jet}. To check the effect of rotation on a monotonically increasing pitch profile we consider equilibria with $P(r) = 1/2 + r^2/2$. The growth rate of the $m=1$, $k=1.0 \; r_a^{-1}$ mode as a function of equilibrium rotation frequency is plotted in Fig. \ref{q_increasing_gamma_vs_rotation} for both force balance models. Similar to the decreasing $P(r)$ cases, rigid rotation is shown to stabilize the kink mode, and we conclude that the rotational stabilization mechanism is not sensitive to the shape of the $P(r)$ profile.

\begin{figure}[t]
\plotone{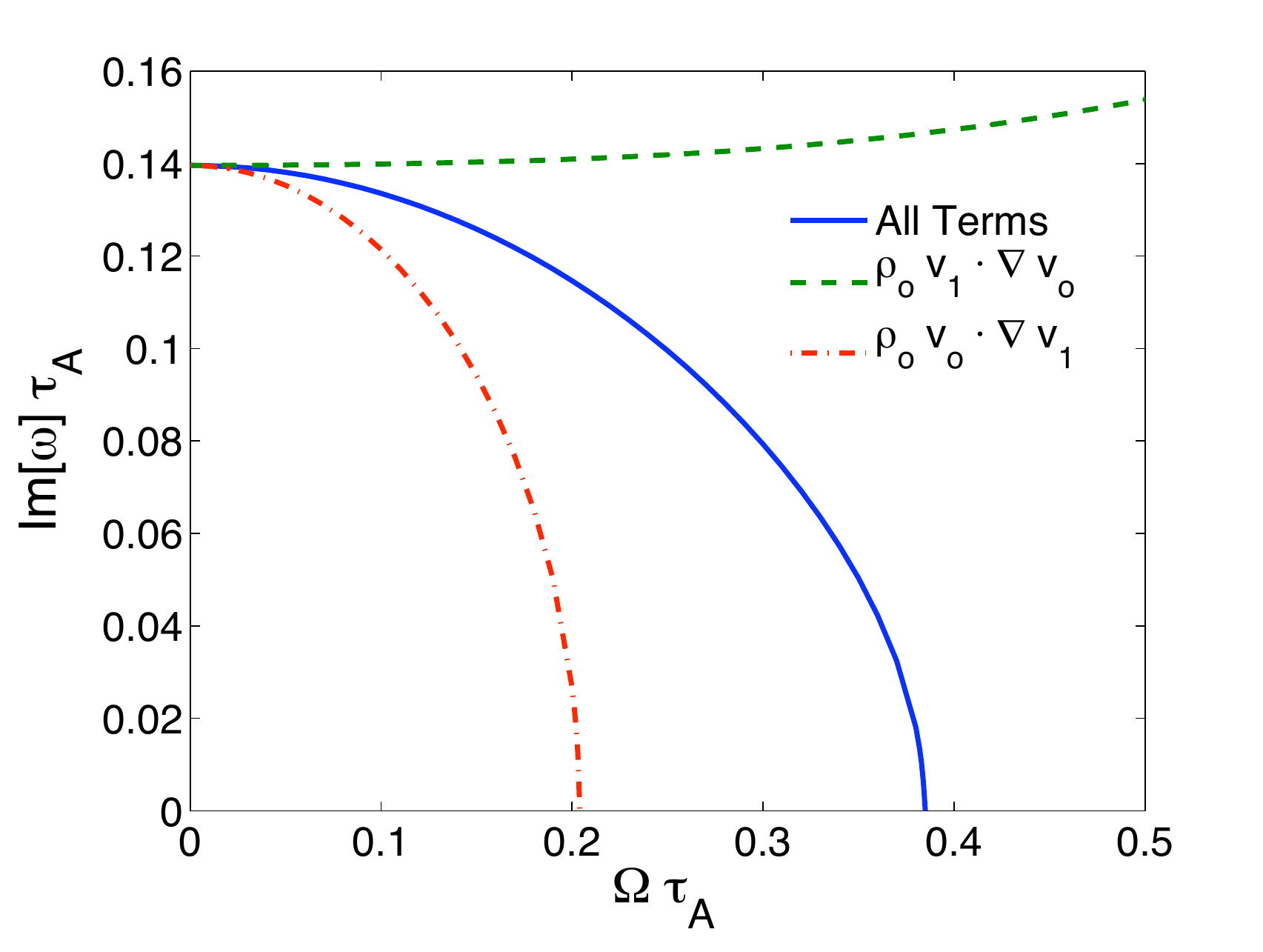}
\caption{Growth rates of the $m=1, k=0.4 \; \lambda_o$ kink mode as a function of rotation frequency with individual inertial terms removed from the linearized momentum equation. For the solid curve all of the terms are present, for the dashed curve the $\rho_o \mathbf v_1 \bm \cdot \bm \nabla \mathbf v_o$ is removed, and for the dot-dashed curve the $\rho_o \mathbf v_o \bm \cdot \bm \nabla \mathbf v_1$ is removed.}
\label{mom_term_removal}
\end{figure}

We also use the eigenmode calculations to investigate the physical mechanism for the rotational stabilization. The linearized momentum equation in the Eulerian frame is given by

\begin{equation}
\frac{\partial \mathbf v}{\partial t} + \rho_0 \mathbf v \bm \cdot \bm \nabla \mathbf v_0 + \rho_0 \mathbf v_0 \bm \cdot \bm \nabla \mathbf v + \rho \mathbf v_0 \bm \cdot \bm \nabla \mathbf v_0 = \mathbf F(\mathbf v),
\label{ler_1}
\end{equation}

\noindent
and the growth rate of the $m=1, k = 0.4 \; \lambda_o$ kink mode is calculated as a function of $\Omega$, removing one equilibrium flow term from the left side at a time. The results are plotted in Fig. \ref{mom_term_removal}, but results with the $\rho \mathbf v_0 \bm \cdot \bm \nabla \mathbf v_0$ term removed are not shown, as this term does not have a significant effect. In the computations without the $\rho_0 \mathbf v \bm \cdot \bm \nabla \mathbf v_0$ term, the growth rate increases with increasing $\Omega$, so this term must play a central role in the stabilization. With rigid rotation, this inertial term is

\begin{equation}
(\mathbf v \bm \cdot \bm \nabla) \mathbf v_0 = -i \; \Omega \; \omega_D (\bm{\hat z} \bm \times \bm \xi) + \Omega^2 (\bm \nabla \bm \cdot \bm \xi) \mathbf r.
\label{ler_2}
\end{equation}

\noindent
By individually removing each of the two terms on the right side of Eq. \ref{ler_2} at a time, we have determined that it is the first term which provides the stabilization. This term contributes to the Coriolis force in the frame of the plasma. 

\begin{figure}[t]
\plotone{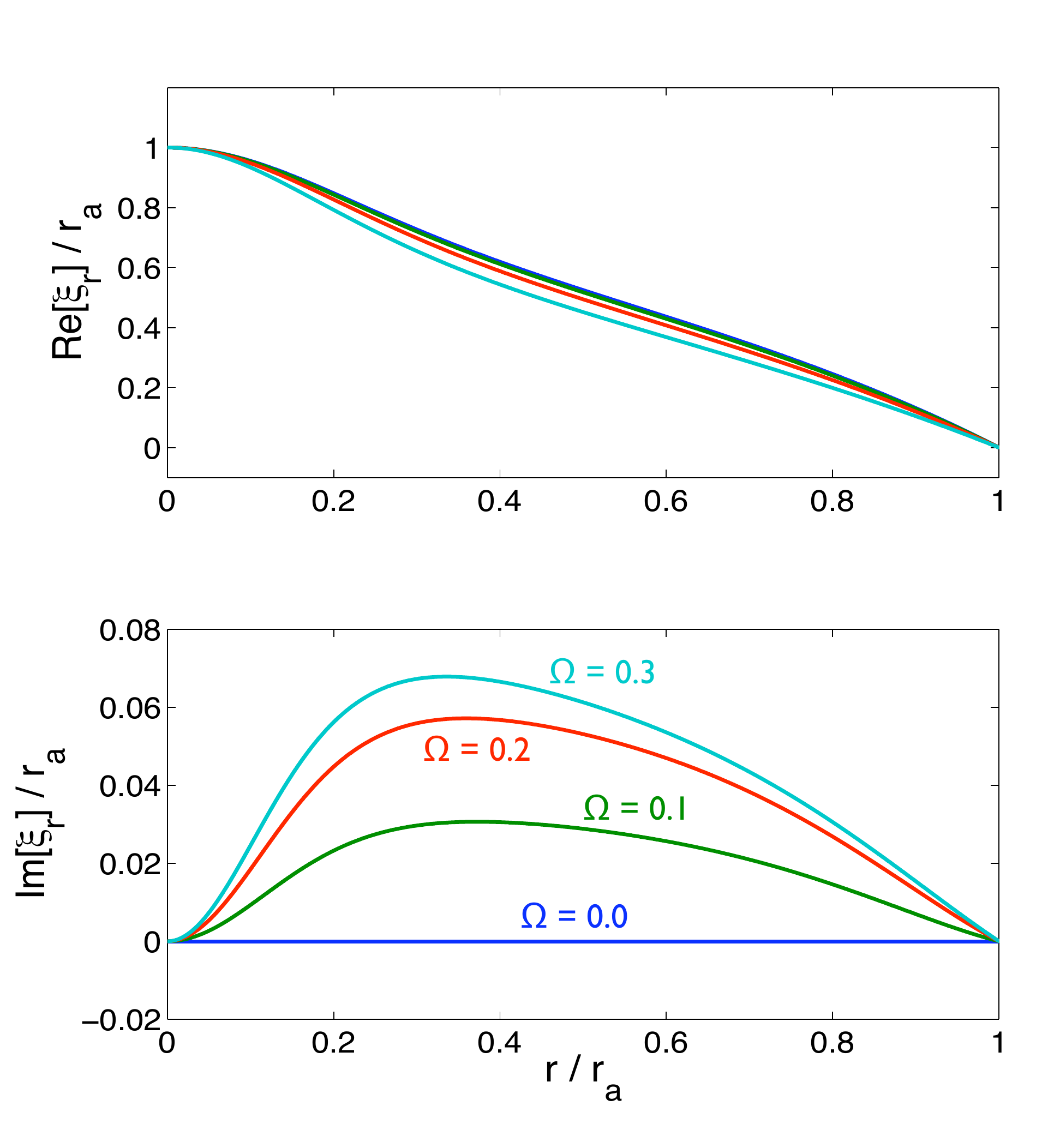}
\caption{The $\xi_r$ component of eigenfunctions of non-resonant $m=1$, $k = 0.4 \: \lambda_o$ kink modes for various equilibrium rotation rates. The eigenmodes are normalized to the maximum value of $\Re[\xi_r]$.}
\label{nonresonant_xi_r}
\end{figure}

\begin{figure}[t]
\plotone{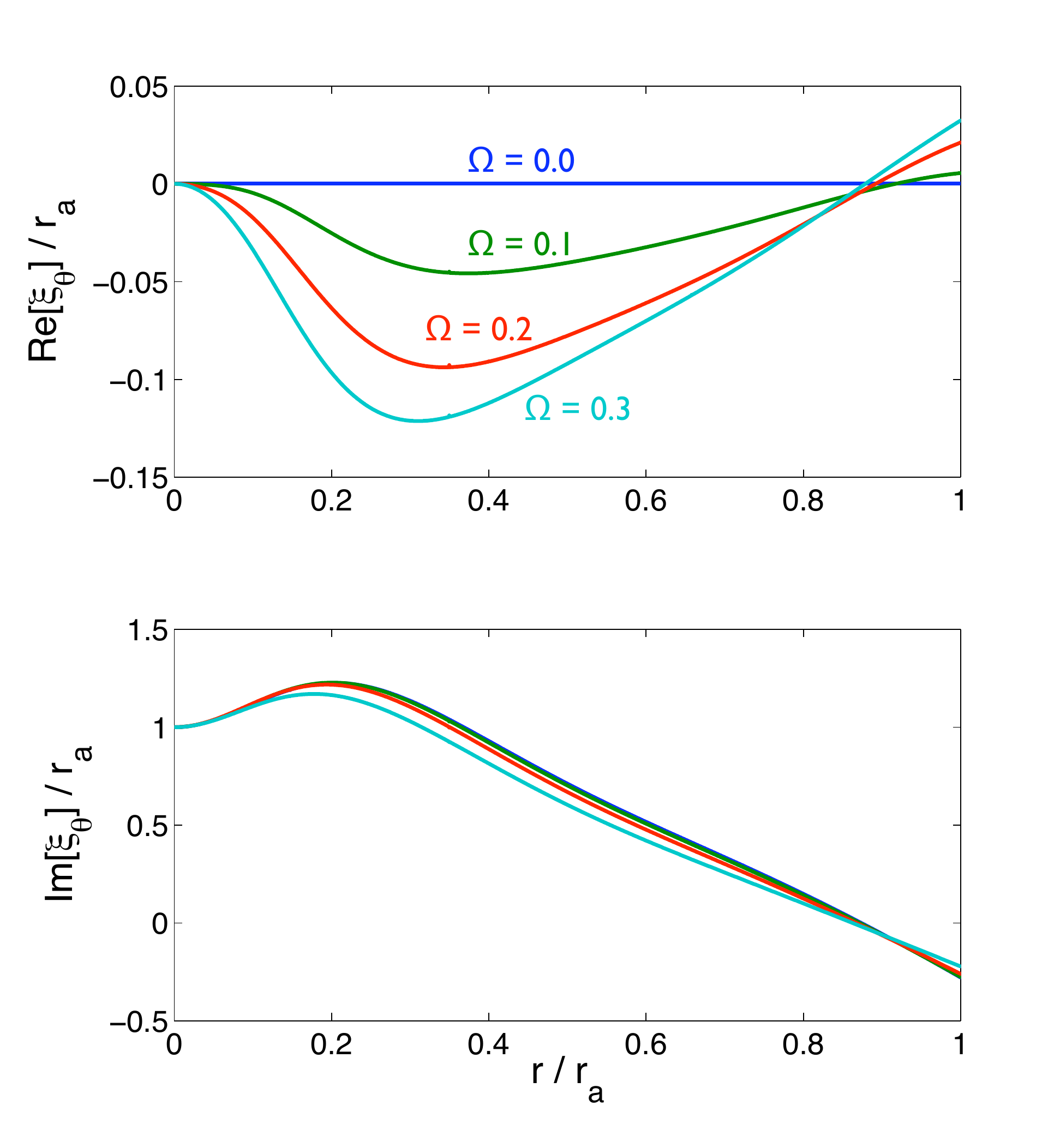}
\caption{The $\xi_\theta$ component of eigenfunctions of non-resonant $m=1$, $k = 0.4 \: \lambda_o$ kink modes for various equilibrium rotation rates. The eigenmodes are normalized to the maximum value of $\Re[\xi_r]$.}
\label{nonresonant_xi_theta}
\end{figure}

Plots of the $\xi_r$ and $\xi_\theta$ components of the eigenfunction for various equilibrium rotation rates are shown in Figs. \ref{nonresonant_xi_r} and \ref{nonresonant_xi_theta}. While there is a slight change in $\Re[\xi_r]$ and $\Im[\xi_\theta]$ as $\Omega$ is varied, the change in $\Im[\xi_r]$ and $\Re[\xi_\theta]$ is more apparent. We note that the Coriolis term locally couples the radial and azimuthal components of translation due to the kink. This distorts the mode giving a radially dependent phase shift in $\xi_r$, and a corresponding change in the real part of $\xi_\theta$. This result is similar to that described in Ref. \citep{Wanex:2005p1603}, where a radially dependent phase shift in the eigenmode due to a sheared equilibrium flow is shown to stabilize the kink mode. Here, we find that a rotational flow without shear introduces a stabilizing distortion of the  mode via the Coriolis force.

It should be noted that the Coriolis term also appears in the $\rho_0 \mathbf v_0 \bm \cdot \bm \nabla \mathbf v$ term in Eq. \ref{ler_1}:

\begin{eqnarray}
(\mathbf v_0 \bm \cdot \bm \nabla) \mathbf v = -i \; \Omega \; \omega_D (\bm{\hat z} \bm \times \bm \xi)  + m \; \Omega \; \omega_D \; \bm \xi \nonumber \\
+ \Omega^2 (\bm \nabla \bm \cdot \bm \xi) \mathbf r - i \; m \; r \; \Omega^2 (\bm \nabla \bm \cdot \bm \xi) \; \bm{\hat \theta},
\label{ler_3}
\end{eqnarray}

\noindent
but when the $(\mathbf v_0 \bm \cdot \bm \nabla) \mathbf v$ term is removed, the stabilization effect is not lost. Equation \ref{ler_3} contains another term which is first order in $\Omega$ given by, $m \; \Omega \; \omega_D \; \bm \xi$. This term provides the Doppler shift in the frequency $\omega$. This Doppler shift appears in the other MHD equations as well. Thus, removing the $\rho_0 \mathbf v_0 \bm \cdot \bm \nabla \mathbf v$ term temporally decouples the velocity field from the magnetic field, reducing the growth rate of the instability, as shown in Fig. \ref{mom_term_removal}.

\begin{figure}[t]
\plotone{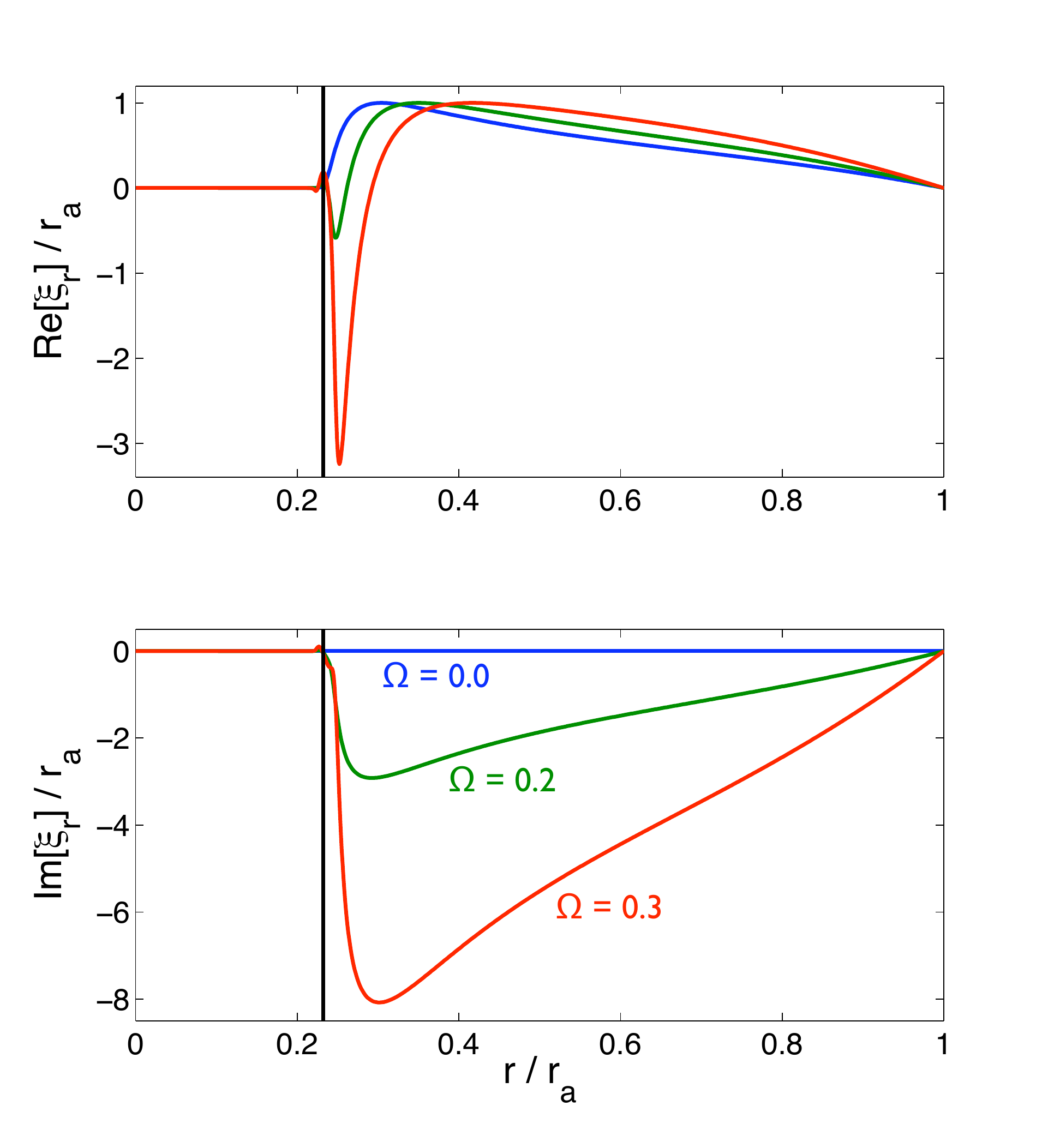}
\caption{The $\xi_r$ component of eigenfunctions of resonant $m=1$, $k = 0.6 \: \lambda_o$ kink modes for various equilibrium rotation rates. The eigenfunctions are normalized to the maximum value of $\Re[\xi_r]$. The vertical black line shows the position of the rational surface at $r = 0.232 \: r_a$.}
\label{resonant_xi_r}
\end{figure}

We also examine the effect of rotation on the resonant eigenmodes. Plots of $\xi_r$ for the $m=1$, $k = 0.6 \: \lambda_o$ kink mode, for various equilibrium rotation rates, are shown in Fig. \ref{resonant_xi_r}. Similar to the non-resonant case, the rotation introduces a significant phase shift in the radial component of the eigenfunction. However, for the resonant case, there is also a significant change in the real part of $\xi_r$ near the rational surface. 

\begin{figure}[t]
\plotone{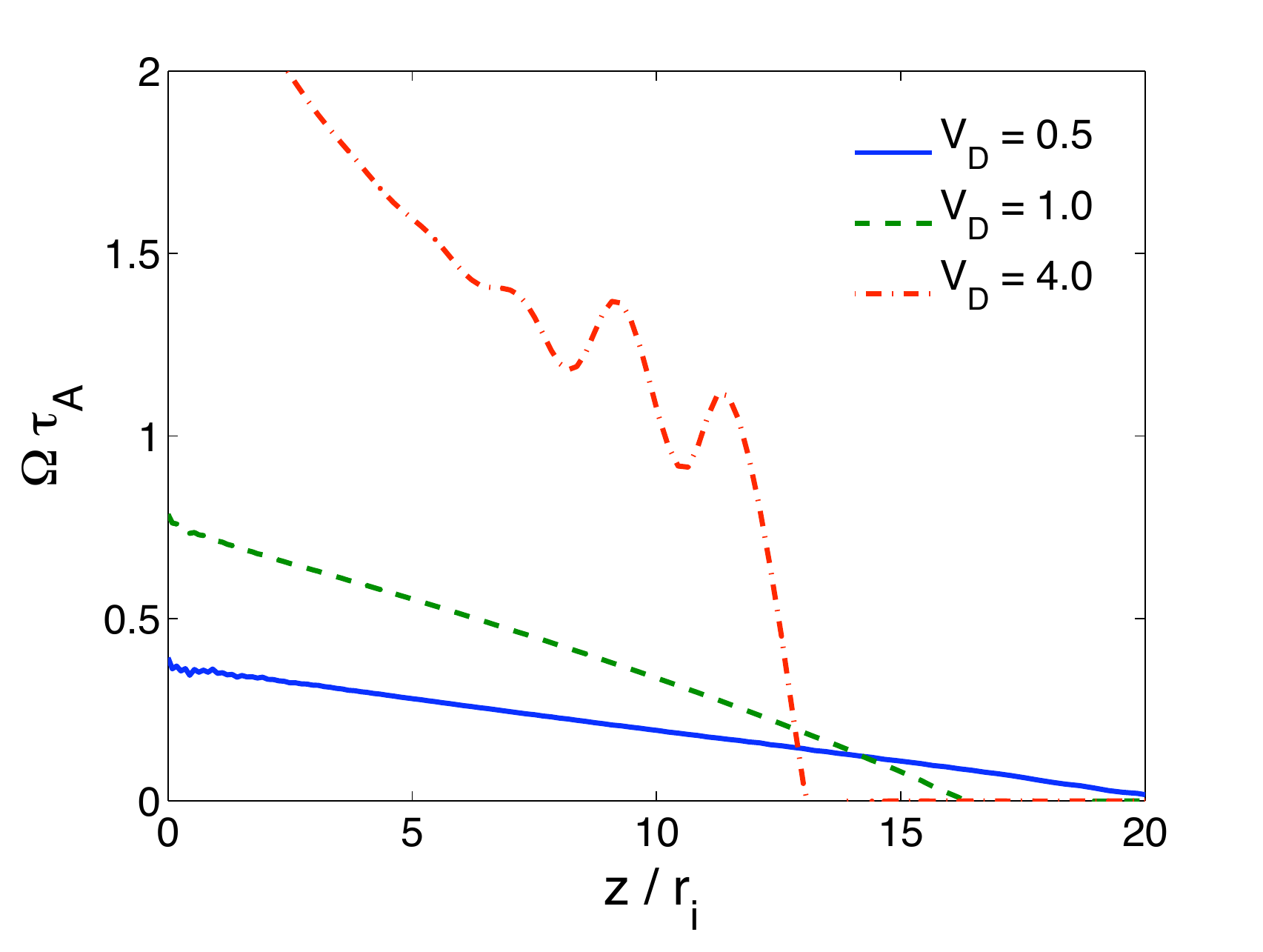}
\caption{Rotation frequency as a function of $z$ in the nonlinear jet calculations discussed in Section \ref{nonlinear_jet}, for various values of $\hat V_D$, at times $t = 8.8, 8.4, 11.3 \; T_i$. Rotation frequency is calculated by making a linear fit to the $m = 0$ component of $v_\theta$ for small values of $r$.}
\label{jet_rot_vs_z}
\end{figure}

To assess the rotation in the simulated magnetic columns described in Sec. \ref{nonlinear_jet}, we calculate the rotation frequencies at different values of $z$. The rotation frequencies plotted in Fig. \ref{jet_rot_vs_z} are determined by making linear fits to the $\theta$-component of the fluid velocity over the radial coordinate. The simulation times chosen for these profiles are such that the kink mode is in the linear phase for the $\hat V_D = 0.5$ and $1.0$ calculations, as can be seen in Fig. \ref{m_1_energy}. It is clear that angular momentum injected by the accretion disk is transported axially by the jet as it expands. As $\hat V_D$ increases, the rotation rate of the jet increases, providing greater stability for the kink mode. 

\begin{figure}[t]
\plotone{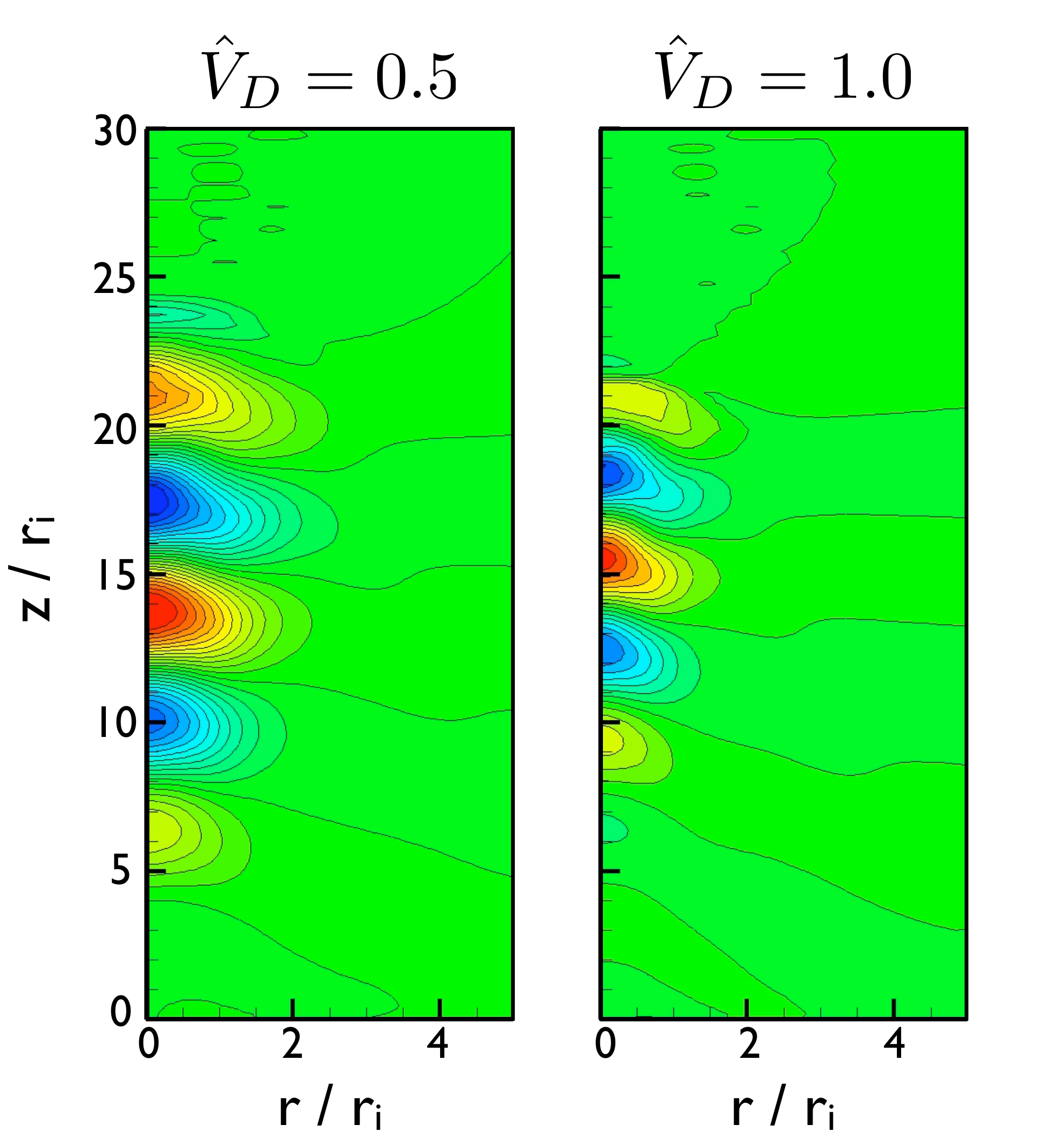}
\caption{The $m=1$ Fourier component of $v_r$ in the nonlinear jet simulations with $\hat V_D = 0.5$ and $1.0$ at times $t = 8.76$ and $10.26 \: T_o$ respectively.}
\label{jet_m_1_vr}
\end{figure}

For comparison to the results of the linear MHD calculations, we examine the $m=1$ kink mode in the nonlinear jet simulations when it is in the linear phase. The $m=1$ Fourier component of $v_r$ is plotted in Fig. \ref{jet_m_1_vr} for the unstable $\hat V_D = 0.5$ and $1.0$ jet simulations at times $t = 8.76$ and $10.26 \: T_i$, respectively. For these times, the kink mode is in its linearly growing phase. Since the modes plotted in Fig. \ref{jet_m_1_vr} extend across the entire width of the jet, we conclude that the kink mode observed in the nonlinear simulations is a non-resonant mode. According to our linear results, these modes would be stable with increased rotation, as is the case in the $\hat V_D = 4.0$ simulation. Similar to the eigenmodes from the linear analysis shown in Fig. \ref{nonresonant_xi_r}, the distortion of the linear eigenmodes in the jet simulations (Fig. \ref{jet_m_1_vr}) is due to a radially dependent phase shift in $v_r$.

\section{Discussion and Conclusions}
\label{conclusions}

Nonlinear non-relativistic MHD simulations of jet evolution, starting from an equilibrium coronal plasma with zero net magnetic flux through the accretion disk, show the formation of a collimated outflow. This outflow is unstable to the current driven $m=1$ kink mode for low rotation velocities of the accretion disk relative to the Alfv\'en speed of the coronal plasma. As it saturates, the kink mode broadens the outflow, but does not destroy the collimation. Similar to previous results \citep{Nakamura:2004p687}, for large rotation velocities of the accretion disk, the outflow is shown to be stable against the kink mode. Moreover, the growth rate of the $m=1$ kink mode is shown to be inversely related to the rotation rate of the accretion disk. This result is counter-intuitive in the sense that as the accretion disk rotates faster, the collimating magnetic field in the jet coils tighter. As the coiling of the magnetic field increases, the current increases. Since the current is the source of free energy for the kink mode, one would expect that the jet would be more unstable for high rotation rates of the accretion disk. However, we observe that it is stable in this regime.

Motivated by the result of the nonlinear jet simulations, we explore the effect of rigid rotation on the $m=1$ kink mode in a periodic cylindrical plasma via linear MHD calculations. The linear calculations are treated as an initial value problem in an Eulerian reference frame and as eigenvalue problems in Eulerian and Lagrangian reference frames. The results from all three methods are in agreement. While previous studies have shown that sheared flow is more efficient at stabilizing the kink mode \citep{Wanex:2005p1603}, we show that rigid equilibrium rotation stabilizes the non-resonant $m=1$ kink mode via the Coriolis effect. The Coriolis effect links radial and azimuthal motions of the plasma, which distorts the kink eigenmode and reduces its growth rate. 

The MHD equations used to model the jet propagation discussed in Section \ref{nonlinear_jet} include dissipative terms, and we should consider what effect dissipation has on the rotational kink stabilization. In order to obtain smooth numerical solutions, the values chosen for the resistivity and the viscosity in the nonlinear jet simulations are much larger than that of any astrophysical jet system. However, we use the dissapationless ideal MHD equations for the eigenvalue analysis discussed in Section \ref{levc}. While dissipation certainly affects the energy densities in the outflow in the jet simulations, the rotational stabilization is an ideal effect and robust to the choice of the dissipation coefficients. 

Our choice of initial conditions in the nonlinear jet simulations discussed in Sec. \ref{nonlinear_jet} has a significant effect on the shape of the magnetic pitch profile, $P(r)$, in the simulated jet. The combination of inertia in the initial coronal plasma and a rapidly decreasing magnetic field acts as a background which the magnetic flux can push against. This allows for the buildup of a large $B_\theta$, producing a monotonically decreasing $P(r)$. In contrast, the simulations of \citet{Moll:2008p4392} produce jets with a monotonically increasing $P(r)$. While these differences affect the shape of the linear eigenfunctions, the eigenvalue calculations discussed in Sec. \ref{linear_eigenvalue_results} show that the rotational stabilization is insensitive to the shape of the $P(r)$ profile. 

With a decreasing $P(r)$ profile and no equilibrium rotation, there are lower and upper bounds on unstable values of $k$ for the kink mode, and the growth rate, $\gamma(k)$, is a function of $k$. This can have a profound effect on the evolution of an expanding jet. The linear rigid rotation calculations discussed in this paper apply only to static equilibria. However, we contend that the results of these calculations can be used as a guide for considering the stability of the time-dependent equilibrium of an expanding jet. As the jet expands, the $k$-value of any given harmonic decreases in time, i.e. the harmonic is stretched by the jet expansion. If we consider an equilibrium that is expanding at a constant rate $s$ with an initial length $L$ and a mode with $k = k'$ at time $t = 0$, the total energy gained by the harmonic over time $t$ is can be estimated as

\begin{equation}
\Delta E(t, k') = E' \int_0^t e^{2 \; \gamma ( \frac{k'}{1 + s t / L}) \; \tilde t} \; d \tilde t ,
\label{ler_0}
\end{equation}

\noindent
where $E'$ is some initial energy in the mode. As we increase $s$, i.e. with faster expansion, the mode spends less time in the unstable range of $k$, and $\Delta E$ decreases. Moreover, equilibrium rotation acts to decrease the area under the $\gamma(k)$ curve, decreasing $\Delta E$ as well. Clearly, this a nonlinear process, and the qualitative description given here motivates further study.

While current driven instabilities may play a role in the wiggled structures which are observed in some outflows \citep{Reipurth:2002p4455, Worrall:2007p3874}; other explanations for these structures have been presented, such as precession of the the source object \citep{Masciadri:2002p4461}. In general, a combination of these effects could contribute to the formation of these structures. Since rotation is shown to stabilize the kink mode, knowledge of the jet rotation velocity relative to the Alfv\'en velocity is critical for understanding the degree to which the kink plays a role. 

\section{Acknowledgements}
\label{acknowledgements}

The authors would like to recognize the following people for their valuable discussions and contributions to this work: John Everett, Ellen Zweibel, Sebastian Heinz, Chris Hegna, Hui Li, and Stirling Colgate. This work is supported by the U.S. Department of Energy Computational Science Graduate Fellowship (DE-FG02-97ER25308) and the National Science Foundation Center for Magnetic Self-Organization in Laboratory and Astrophysical Plasmas (PHY 0821899). Nonlinear simulations were performed at the National Energy Research Scientific Computing Center, which is supported by the Office of Science of the U.S. Department of Energy under Contract No. DE-AC02-05CH11231.


\end{document}